\documentclass[aps,pra,twocolumn,showpacs,groupedaddress,floatfix,amsmath,amssymb]{revtex4}
\usepackage{graphicx}

\begin{document}
\title{Rapidly rotating boson molecules with long or short range 
repulsion: \\ an exact diagonalization study}
\author{Leslie O. Baksmaty, Constantine Yannouleas, and Uzi Landman}
\affiliation{School of Physics, 
Georgia Institute of Technology, Atlanta, Georgia 30332-0430, USA}
\date{31 October 2006}

\begin{abstract}
The Hamiltonian for a small number, $N \leq 11$, of bosons in a rapidly rotating
harmonic trap, interacting via a short range (contact potential) or a long range
(Coulomb) interaction, is studied via an exact diagonalization in the lowest 
Landau level. Our analysis shows that, for both low and high fractional fillings,
the bosons localize and form rotating boson molecules (RBMs) consisting of 
concentric polygonal rings. Focusing on systems with the number of trapped atoms
sufficiently large to form multi-ring bosonic molecules, we find that, as a 
function of the rotational frequency and regardless of the type of repulsive 
interaction, the ground-state angular momenta grow in specific steps that 
coincide with the number of localized bosons on each concentric ring. Comparison
of the conditional probability distributions (CPDs) for both interactions 
suggests that the degree of crystalline correlations appears to depend more 
on the fractional filling $\nu$ than on the range of the interaction. The RBMs 
behave as nonrigid rotors, i.e., the concentric rings rotate independently of 
each other. At filling fractions $\nu < 1/2$, we observe well developed 
crystallinity in the CPDs (two-point correlation functions). For larger filling 
fractions $\nu > 1/2$, observation of similar molecular patterns requires 
consideration of even higher-order correlation functions.
\end{abstract}
\maketitle
\section{Introduction}

Rotating ultra-cold trapped Bose condensed systems are most commonly discussed
in the context of formation of vortex lattices, which are solutions to the
Gross-Pitaevskii (GP) mean-field equation 
\cite{stri,legg,ho,fett,peth,pita,baks1,baks2,peth2,baym1}. Such vortex lattices
have indeed been found experimentally for systems containing a large number of 
bosons \cite{dali,kett,corn}.
Nevertheless, several theoretical investigations 
\cite{wilk1,vief,wilk2,regn1,regn2} of 
{\it rapidly\/} rotating trapped bosonic systems suggested formation of strongly
correlated exotic states which differ drastically from the aforementioned 
vortex-lattice states. While experimental realizations of such strongly 
correlated states have not been reported yet, there is already a significant 
effort associated with two-dimenional (2D) exact-diagonalization (EXD) studies 
of a small number of particles ($N$) in the lowest Landau level, LLL; the LLL 
restriction corresponds to the regime of rapid rotation, where the rotational
frequency of the trap $\Omega$ equals the frequency $\omega_0$ of the confining 
potential. The large majority \cite{wilk1,wilk2,regn1,regn2} of such EXD studies 
have attempted to establish a close connection between rapidly rotating bosonic 
gases and the physics of elecrtons under fractional-quantum-Hall-effect (FQHE) 
conditions employing the bosonic version of ``quantum-liquid'' analytic wave 
functions, such as the Laughlin wave functions, composite-fermion, Moore-Read, 
and Pfaffian functions.

Recently, the ``quantum-liquid'' picture for a small number of trapped
electrons in the FQHE regime has been challenged in a series of extensive 
studies \cite{yl1,yl2,yl3,yl4,yl5,yl6} of electrons in 2D quantum dots under 
high magnetic fields ($B$). Such studies (both EXD and variational) revealed 
that, at least for finite systems, the underlying physical picture governing
the behavior of strongly-correlated electrons is not that of a ``quantum 
liquid.'' Instead, the appropriate description is in terms of a ``quantum 
crystal,'' with the localized electrons arranged in polygonal concentric rings
\cite{yl1,yl2,yl3,yl4,yl5,bao,seki,maks}. 
These ``crystalline'' states lack \cite{yl3,yl5}  the familiar rigidity of a 
classical extended crystal, and are better described
\cite{yl1,yl2,yl3,yl4,yl5,yl6} as {\it rotating electron (\/}or {\it Wigner) 
molecules\/} (REMs or RWMs).   

Motivated by the discovery in the case of electrons of REMs at high $B$ (and 
from the fact that Wigner molecules form also at zero magnetic field
\cite{yl7,egge,yl8,mikh,yl9,yl10}) 
some theoretical studies have most recently shown that analogous 
molecular patterns of localized bosons do form in the case of a small number of 
particles inside a static or rotating harmonic trap
\cite{yl11,yl12,yl13,mann,barb}. In analogy with the electron case, the bosonic 
molecular structures can be referred to \cite{yl13} as {\it rotating boson 
molecules\/} (RBMs); a description of RBMs via a variational wave function
\cite{note1} built from symmetry-breaking displaced Gaussian orbitals with 
subsequent restoration of the rotational symmetry was presented in Refs.\ 
\cite{yl11,yl12,yl13}. 

In this paper, we use exact diagonalization in the 
lowest Landau level to investigate the 
formation and properties of RBMs focusing on a larger number of particles than
previously studied, in particular for sizes where multiple-ring formation 
can be expected based on our knowledge of the case of 2D electrons in high $B$.
We study a finite number of particles ($N \le 11$) at {\it both\/} low 
($ \nu < 1/2$) and high ($ \nu \geq 1/2$) filling fractions
$\nu \equiv N(N-1)/2L$ \cite{note56} (where $L \equiv {\cal L}/\hbar$ is the 
quantum number associated with the total angular momentum ${\cal L}$) and 
investigate both the cases of a long (Coulomb) and a short ($\delta$-function) 
range repulsive interaction. 

As in the case of electrons in 2D quantum dots, we
probe the crystalline nature of the bosonic ground states by calculating the full
anisotropic two-point correlation function $P({\bf r},{\bf r}_{0})$ 
associated with the exact wavefunction $\Psi({\bf r}_1,{\bf r}_2,...,{\bf r}_N)$.
The quantity $P({\bf r},{\bf r}_{0})$ is proportional to the probability of 
finding a boson at ${\bf r}$ given that there is another boson at the 
observation point ${\bf r}_{0}$, and it is often referred to as the conditional 
probability distribution \cite{yl4,yl5,yl8,yl10,yl11,yl13} (CPD).
A main finding of our study is that consideration solely of the CPDs is not
sufficient for the boson case at high fractional fillings $\nu \geq 1/2$;
in this case, one needs to calculate even higher-order correlation functions,
e.g., the full $N$-point correlation function 
$P( {\bf r}; {\bf r}_1,{\bf r}_2,...,{\bf r}_{N-1} )$ [see Eq.\ 
(\ref{npoint_def})]. 

The present investigations are also motivated by recent experimental 
developments, e.g., the emergence of trapped ultra-cold gas assemblies featuring
bosons interacting via a long-range dipole-dipole interaction. 
We expect the results of this paper to be directly relevant to such systems with
a two-body repulsion intermediate between the Coulomb and the delta potentials.
Additionally, we note the appearance of promising experimental techniques for 
measuring higher-order correlations in ultra-cold gases employing an atomic 
Hanbury Brown-Twiss scheme \cite{orsay} or shot-noise 
interferometry \cite{shot_noise_1,shot_noise_2}.
Experimental realization of few-boson rotating systems can be anticipated in
the near future as a result of increasing sophistication of experiments 
involving periodic optical lattices co-rotating with 
the gas, which are capable of holding a few atoms in each site. 
A natural first step in the study of such systems is the analysis of the 
physical properties of a few particles confined in a rotating trap with open 
boundary conditions (i.e., conservation of the total angular momentum $L$), 
such as we do here. 

\begin{figure}[t]
\centering
\includegraphics[width=8cm]{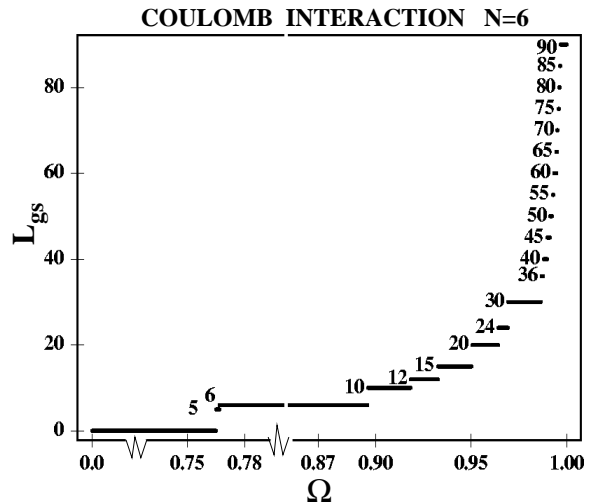}
\caption{Ground-state angular momenta, $L_{\text{gs}}$, for $N=6$ bosons in a 
rapidly rotating trap [described by the LLL Hamiltonian in Eq.\ (\ref{H_lll})], 
as a function of the rotational frequency $\Omega$ expressed in
units of $\omega_\bot$. The 
bosons interact via a Coulombic repulsion and the many-body Hilbert space is 
restricted to the LLL. The angular momentum associated with the first bosonic 
Laughlin state occurs at $L=30$, i.e., at $N(N-1)$. The value of 
$c=0.2 \hbar \omega_\bot \Lambda$ for
this plot; the many-body wave function does not depend on this 
choice. Note the stepwise variation of the values 
of the ground-state angular momenta, indicating the presence of
an {\it intrinsic\/} point-group symmetry associated with the polygonal-ring
structure of RBMs (see text for details).}
\label{6_m_vs_om}
\end{figure}

Our main results can be summarized as follows: 
Similar to the well-established (see, e.g., Refs.\ \cite{yl4,yl5}) formation of 
rotating electron molecules in quantum dots, we describe here the emergence of 
rotating boson molecules in rotating traps. The RBMs are also organized in 
concentric polygonal rings that rotate independently of each other, and the 
polygonal rings correspond to classical equilibrium configurations and/or their 
low-energy isomers. Furthermore, the degree of crystallinity increases gradually
with larger angular momenta $L$'s (smaller filling fractions $\nu$'s), as was 
the trend \cite{yl2,yl4,yl5} for the REMs and as was observed also for 
$\nu < 1/2$ in another study \cite{mann} for rotating bosons in the LLL with 
smaller $N$ and single-ring structures. We finally note that the crystalline 
character of the RBMs appears to depend only weakly on the range of the 
repelling interaction, for both the low (see also Ref.\ \cite{mann}) and high 
(unlike Ref.\ \cite{barb}) fractional fillings \cite{note2}.

In studies of 2D quantum dots, CPDs were used some time ago in Refs.\ 
\cite{seki,maks,yl8,note67}. For probing the intrinsic molecular structure 
in the case of ultracold bosons in 2D traps, however, they were introduced only
recently by Romanovsky {\it et al.\/} \cite{yl11}. The importance of using CPDs 
as a probe can hardly be underestimated. Indeed, while EXD calculations  
for {\it bosons\/} in the LLL have been reported earlier
\cite{wilk1,vief,wilk2,regn1,regn2}, the analysis in these studies did not
include calculations of the CPDs, and consequently formation of 
rotating boson molecules was not recognized.

This paper is organized as follows. In section~\ref{Methods} we 
establish our notation and detail the physical assumptions that underlie
the construction of our {\it ab initio\/} Hamiltonian. 
In section~\ref{results}, we present our results for $N=6,9$ and 11 bosons 
and compare the emerging crystallinity for both Coulomb and $\delta$-function 
interactions. We summarize our results in section IV. 
The appendix presents explicit formulas for calculations of  
two-body matrix elements between two permanents. These permanents
are used as the basis wave functions that span the many-boson
Hilbert space; for the case of fermions, of course, one uses the more
familiar Slater determinants \cite{yl2}.
 
\begin{figure}[t]
\centering
\includegraphics[width=8cm]{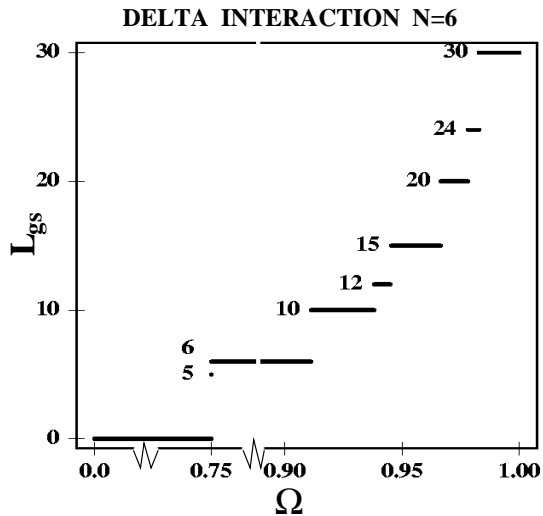}
\caption{Ground-state angular momenta, $L_{\text{gs}}$, for $N=6$ bosons in a 
rapidly rotating trap [described by the LLL Hamiltonian in Eq.\ (\ref{H_lll})], 
as a function of the rotational frequency $\Omega$ expressed in units of
$\omega_\bot$. The 
bosons interact via a {\it delta\/} repulsion and the many-body Hilbert space is 
restricted to the LLL. The angular momentum associated with the first bosonic 
Laughlin state occurs at $L=30$, i.e., at $N(N-1)$. The value of 
$g=2\pi \hbar \omega_\bot \Lambda^2/N$ for this plot; 
the many-body wave function does not 
depend on this choice. The values of the angular momenta terminate with the 
value $L=30$ (the Laughlin value) at $\Omega/\omega_\bot=1$. In 
contrast, in the Coulomb-interaction case (see Fig.\ \ref{6_m_vs_om}), the values
of the ground-state angular momenta do not terminate, but diverge as 
$\Omega/\omega_\bot \rightarrow 1$. Note the stepwise variation of the values 
of the ground-state angular momenta in both cases, indicating the presence of
an {\it intrinsic\/} point-group symmetry associated with the polygonal-ring
structure of RBMs (see text for details).}
\label{6_m_vs_om_dlt}
\end{figure}

\section{Theoretical methods}
\label{Methods}
\subsection{Ab initio Hamiltonian in the LLL}

Requiring a very strong confinement of the harmonic trap 
along the axis of rotation ($\hbar \omega_{z} >> \hbar \omega_{\bot}$), 
freezes out the many body dynamics in the $z$-dimension, and the wavefunction
along this direction can be assumed to be permanently in the corresponding 
oscillator ground state.  
We are thus left with an effectively 2D system. For such a setup, 
the Hamiltonian for $N$ atoms of mass $m$ in a harmonic trap $(\omega_\bot)$ 
rotating at angular frequency $\Omega \hat{\bf z}$ is given by:  
\begin{equation}
H=\sum_{i=1}^N
\left( \frac{{\bf p}_{i}^{2}}{2m}+\frac{1}{2} 
m \omega_{\bot}^{2} {\bf r}_{i}^{2} \right)
+\sum_{i < j}^{N} v({\bf r}_{i}-{\bf r}_{j})
-\Omega {\cal L}.
\label{3d_hamiltonian_1}
\end{equation}
Here ${\cal L}=-\hbar L=\sum_{i=1}^N \hat{{\bf z}}\!\cdot{\bf r}_i 
\times {\bf p}_{i}$ is the total angular momentum; $ {\bf r } \; =\; (x,y)$ and
$ {\bf p } \; =( p_x,p_y )$ represent the single-particle position and 
angular momentum in the $x-y$ plane and $\omega_{\bot}$ is the radial trap 
frequency.

\begin{figure*}
\centering
\includegraphics*[width=15cm]{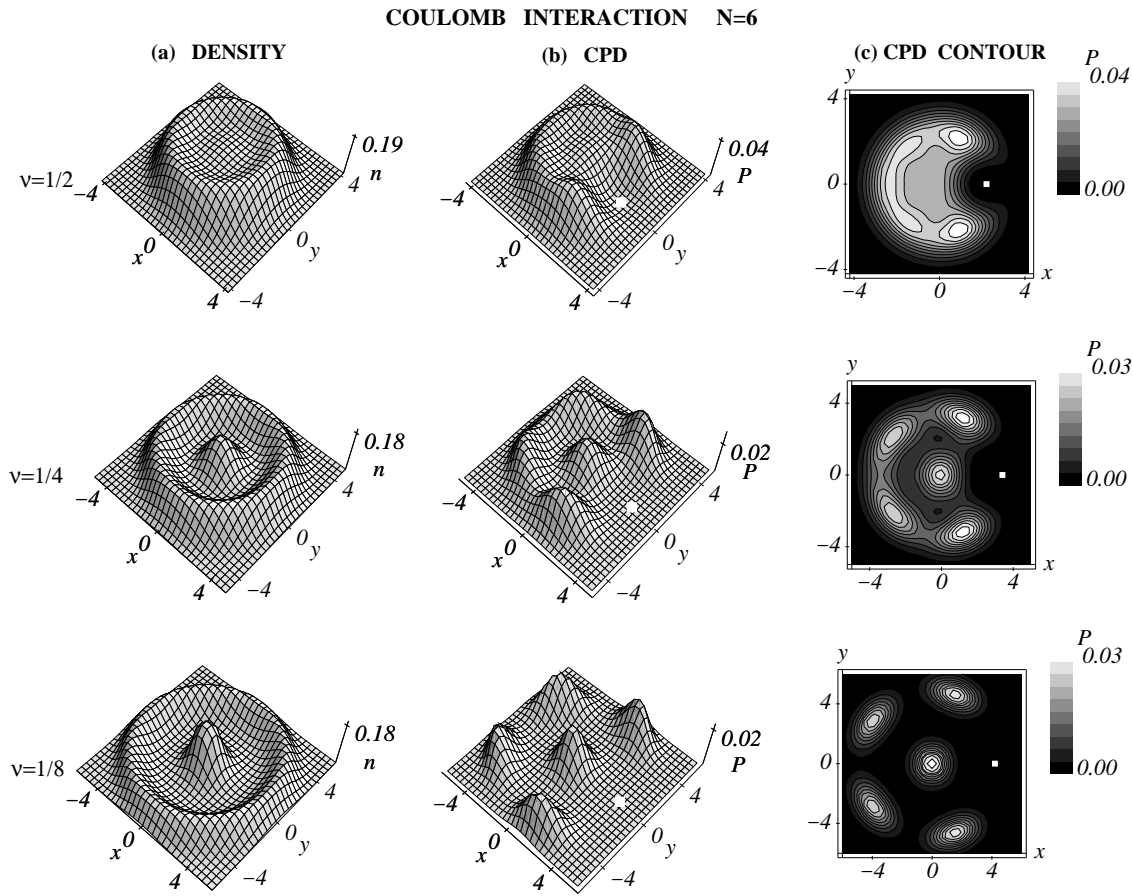}
\caption{(a) Single-particle densities [$n({\bf r})$; left column], 
(b) CPDs $\left[ P({\bf r},{\bf r}_{0}) \right]$ in 3D plots (middle column),
and (c) CPDs in contour plots (right column), portraying the strengthening 
of the crystalline RBM 
structure for $N=6$ bosons interacting via a repulsive Coulomb interaction as 
the filling fraction $\nu$ is reduced. The white dots in the CPD 
plots indicate the reference point 
${\bf r}_0$. We note in particular the gradual enhancement of the peak at the 
center of the plots, and the growth of the radius of the outer ring; the latter
reflects the nonrigid-rotor nature of the RBMs (in analogy with the findings
of Ref.\ \cite{yl5} regarding the properties of rotating electron
molecules). The cases of $\nu=1/4$ and $\nu=1/8$ exhibit a clear $(1,5)$
crystalline arrangement, while the case of $\nu=1/2$ (first Laughlin state) is 
intermediate between a $(1,5)$ and a $(0,6)$ pattern (see text for details).
Lengths in units of $\Lambda$. The vertical scales are in arbitrary 
units, which however do not change for the panels within the same 
column (a), (b), or (c).}
\label{6_coulomb_crystallization}
\end{figure*}

The Hamiltonian can be rewritten in the form,
\begin{eqnarray}
H&=&\sum_{i=1}^N \left\{ \frac{\left({\bf p}_{i}-m\Omega\hat{\bf z} \times 
{\bf r}_{i}\right)^{2}}{2 m}+\frac{m}{2}(\omega_{\bot}^2-\Omega^2) {\bf r}_i^2
\right\} \nonumber \\ 
&&~~~~~~~~~~~+\sum_{i < j}^{N} v({\bf r}_{i}-{\bf r}_{j}). 
\label{3d_hamiltonian_3}
\end{eqnarray}
The kinetic part of this Hamiltonian is formally equivalent to that of the 
Hamiltonian in the symmetric gauge of an electron (of charge $e$ and mass
$m_e$) moving in two dimensions under a constant perpendicular magnetic field 
$B$, if one makes the identification that the cyclotron frequency
$\omega_c=eB/(m_ec)\rightarrow 2\Omega$.

We are now ready to examine our main approximation which is based on 
two assumptions: (1) that the rotational frequency $\Omega$ is close to
that of the confining trap; in this case the external confinement can be
neglected in a first approximation and thus the single-particle spectra are
organized in infinitely-degenerate Landau levels that are 
separated by an energy gap of $2\hbar \omega_\bot$, and (2)  
that the interaction strength is so weak that the mixing of 
Landau levels can be ignored. Since we work at zero temperature it then follows 
that all particles are in the lowest Landau level. This physical regime 
is known in the literature as the weak interaction limit. In a recent 
paper \cite{feder} it has been verified by explicit calculation that this 
approximation, i.e., the lowest Landau level approximation, is very good for 
$\nu < 1/2$ (low fractional fillings), where strong crystalline correlations are
most developed, as reported in the present work.

We note, however, that strong correlations beyond the Gross-Pitaevskii mean field
persist in the LLL even for $\nu > 1/2$ (high fractional fillings), as was found 
via EXD calculations for smaller $N$ in Ref.\ \cite{barb}. In particular, Ref.\ 
\cite{barb} concluded that even in the most favorable case (i.e., 
for $L=N$) formation of vortices for small $N$ is not a prevalent phenomenon,
and their appearance ``is restricted to the vicinity of some critical values
of the rotational frequency $\Omega$.'' This conclusion concurs with the
results of Ref.\ \cite{yl13} which used RBM variational wave functions to
study a broad range of rotational frequencies that do not lead to the 
restriction to the LLL. The present work brings new insights to these
findings by showing that the crystalline correlations for $\nu > 1/2$ (high
fractional fillings) are clearly seen in the $N$-point correlation functions, 
even though an inspection of the CPDs alone may be inconclusive 
(see examples in Section III.A).

Taking into account that the single-particle spectrum associated with the
Hamiltonian (\ref{3d_hamiltonian_3}) (Fock-Darwin spectrum \cite{fock,darw}) 
is given by 
$\epsilon^{\text{FD}}_{n,l}=\hbar [(2n+|l|+1) \omega_\bot - l \Omega]$, the 
restriction to the LLL requires $n=0$ (Fock-Darwin single-particle states with 
zero radial nodes) and reduces the Hamiltonian $H$ above 
[Eq.\ (\ref{3d_hamiltonian_3})] to the simpler form: 
\begin{equation}
H_{\text{LLL}}=N \hbar \omega_\bot + \hbar(\omega_{\bot}-\Omega) L
+\sum_{i < j}^{N} v({\bf r}_{i}-{\bf r}_{j}),
\label{H_lll}
\end{equation}
for which only the interaction term is non-trivial, since the many-body energy 
eigenstates are eigenstates of the total angular momentum as well; 
$L=\sum_{i=1}^N l_i$.

In this study we shall investigate and compare our results for strongly
correlated states in the LLL obtained for a system of bosons interacting via a 
long or short range force respectively represented 
by the coulomb interaction,
\begin{equation}
v({\bf r})=\frac{c}{|{\bf r}|}, 
\end{equation}
or the contact potential,
\begin{equation}
v({\bf r})=g\delta^2({\bf r}). 
\label{interactions}
\end{equation}

As may be inferred from the form of $H_{\text{LLL}}$ the coupling constants  
$c$ and $g$ trivially change the result by a multiplicative factor 
and their importance occurs only in comparing our results to a particular 
experimental system. Mathematically expressed, our description is valid 
only when:  
\begin{equation}
g \Lambda^{-2} \ll 2\hbar \omega_{\bot} \ll \hbar \omega_{z}
\label{delta_weak}
\end{equation}
in the case of a contact potential, and 
\begin{equation}
\frac{c} {\Lambda} \ll 2 \hbar \omega_{\bot} \ll \hbar \omega_{z},
\label{coulomb_weak}
\end{equation}
for the coulomb interaction, with
$\Lambda = \sqrt{\hbar/(m\omega_\bot)}$ being the characteristic length
of the harmonic trap.  In these inequalities we have, for the sake of 
completeness, also included the energy scale associated with the 'frozen' 
perpendicular direction to give a full picture of the hierarchy of energy 
scales implicit in our approximations. 

\begin{figure*}[t]
\centering
\includegraphics*[width=15cm]{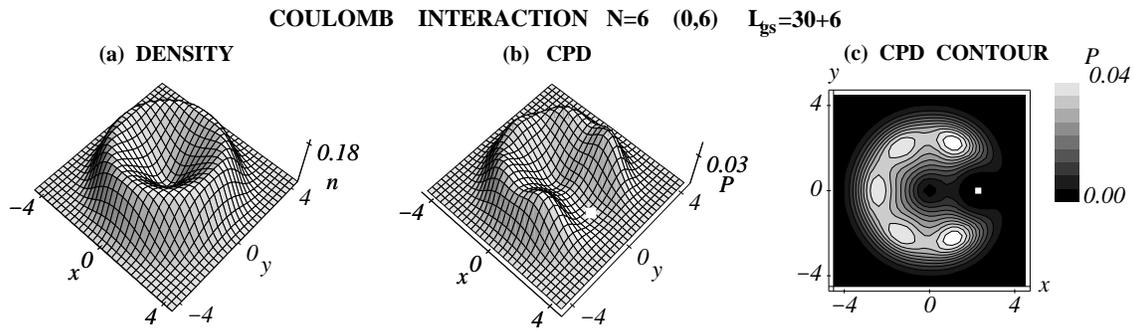}
\caption{A ground state with a well-developed $(0,6)$ arrangement for $N=6$ 
bosons interacting via a Coulomb repulsion at angular momentum 
$L_{\text{gs}}=36$. In classical electrostatic 
calculations \cite{peeters_classical}, the $(0,6)$ arrangement is the first
metastable isomer to the most stable $(1,5)$ one shown by the CPDs of 
Fig.\ \ref{6_coulomb_crystallization}. The individual panels are: (a) $3D$ plot 
of the single-particle density $n({\bf r})$; (b) $3D$ plot of the CPD; and 
(c) Contour plot of the CPD. In the CPD $\left[ P({\bf r},{\bf r}_{0}) \right]$ 
plots, the white dots indicate the reference point ${\bf r}_0$.
Lengths in units of $\Lambda$. In (b) and (c), the vertical scales
are the same.}
\label{6_isomer}
\end{figure*}

\begin{figure}[b]
\centering
\includegraphics*[width=7.0cm]{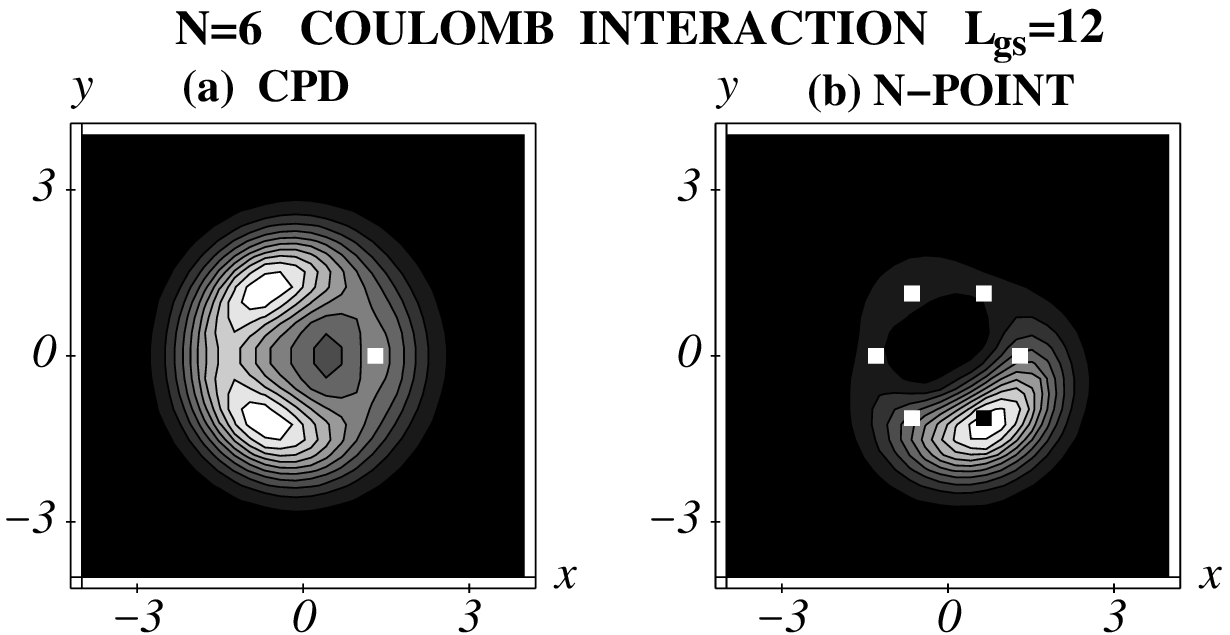}
\caption{Contour plots of the CPD (a) and $N$-point correlation function (b) for
$N=6$ bosons with $L_{\text{gs}}=12$ interacting via a Coulomb repulsion. 
The white squares indicate the positions of the fixed particles. The black 
square in (b) indicates the position of the 6th particle according to the 
classical $(0,6)$ molecular configuration. Note that the CPD in (a) fails to
reveal the $(0,6)$ pattern, which, however, is clearly seen in the $N$-point
correlation function in (b). Lengths in units of $\Lambda$. 
The vertical scales are arbitrary.}
\label{6_12_npoint_clmb}
\end{figure}

\begin{figure}[b]
\centering
\includegraphics*[width=7.0cm]{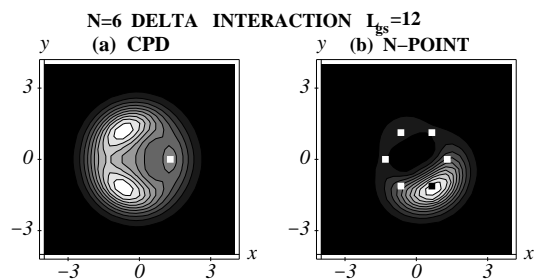}
\caption{Contour plots of the CPD (a) and $N$-point correlation function (b) for
$N=6$ bosons with $L_{\text{gs}}=12$ interacting via a $\delta$-repulsion.
The white squares indicate the positions of the fixed particles. The black
square in (b) indicates the position of the 6th particle according to the
classical $(0,6)$ molecular configuration. Note that the CPD in (a) fails to
reveal the $(0,6)$ pattern, which, however, is clearly seen in the $N$-point
correlation function in (b). Lengths in units of $\Lambda$. The 
vertical scales are arbitrary.}
\label{6_12_npoint_delta}
\end{figure}

\subsection{EXD wave function and Conditional Probability Distribution}
\label{cpd_section}

\begin{figure*}[t]
\centering
\includegraphics*[width=14.0cm]{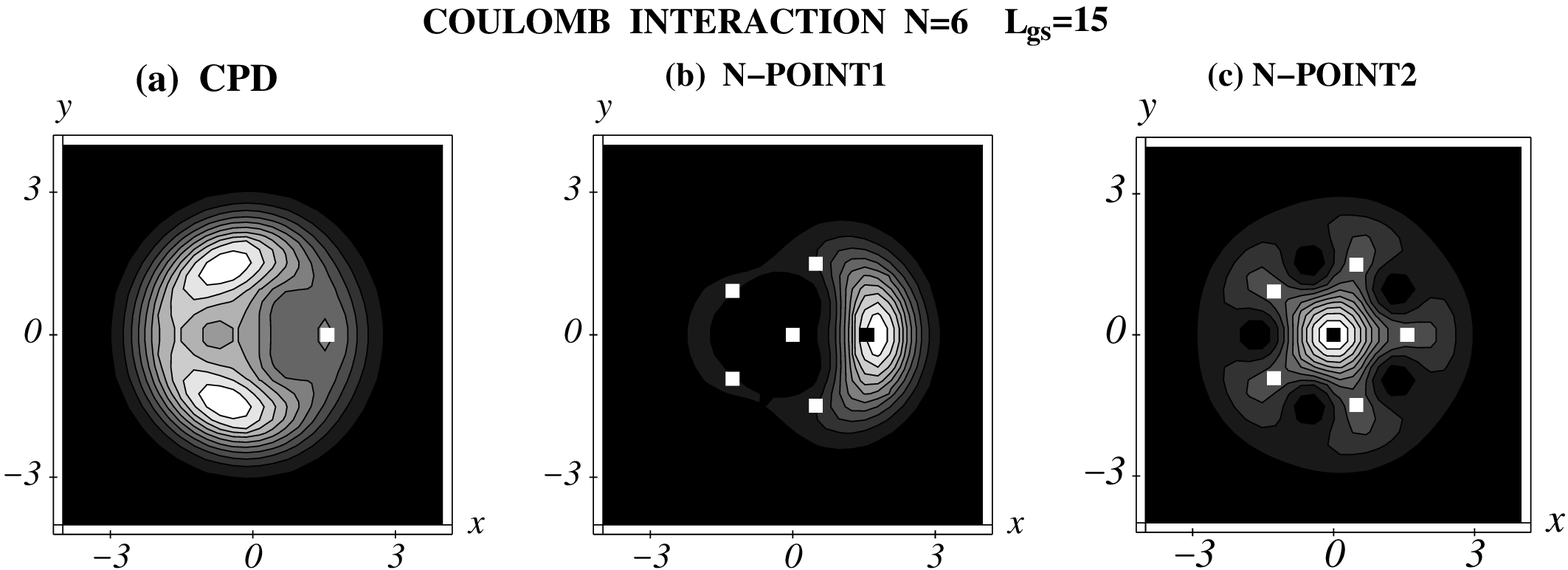}
\caption{Contour plots of the CPD (a) and $N$-point correlation function (b) and
(c) for $N=6$ bosons with $L_{\text{gs}}=15$ interacting via a Coulomb 
repulsion. The white squares indicate the positions of the fixed particles. 
The black square in (b) and (c) indicates the position of the 6th particle 
according to the classical $(1,5)$ molecular configuration. 
Note the different arrangements of the five fixed particles, i.e., (b) one fixed
particle at the center and (c) no fixed particle at the center. Note also that 
the CPD in (a) fails to reveal the $(1,5)$ pattern, which, however, is clearly 
seen in the $N$-point correlation functions in both (b) and (c).
Lengths in units of $\Lambda$. The vertical scales are arbitrary, but
the same in (b) and (c).}
\label{6_15_npoint_clmb}
\end{figure*}

For the solution of the large scale, but sparse, matrix eigenvalue problem
associated with the Hamiltonina $H_{\text{LLL}}$,
we have used the ARPACK computer code \cite{arp,petsc}. For a given $L$, 
the EXD many-body wave function is a linear superposition of permanents
made out of the single-particle LLL wave functions,

\begin{figure*}[t]
\centering
\includegraphics*[width=14.0cm]{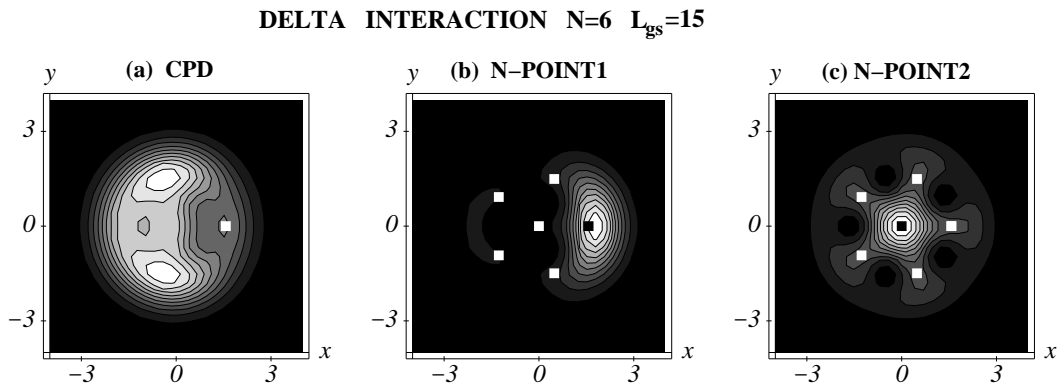}
\caption{Contour plots of the CPD (a) and $N$-point correlation function (b) and
(c) for $N=6$ bosons with $L_{\text{gs}}=15$ interacting via a 
$\delta$-repulsion. The white squares indicate the positions of the fixed 
particles. The black square in (b) and (c) indicates the position of the 6th 
particle according to the classical $(1,5)$ molecular configuration. 
Note the different arrangements of the five fixed particles, i.e., (b) one fixed
particle at the center and (c) no fixed particle at the center. Note also that 
the CPD in (a) fails to reveal the $(1,5)$ pattern, which, however, is clearly 
seen in the $N$-point correlation functions in both (b) and (c).
Lengths in units of $\Lambda$. The vertical scales are arbitrary, but
the same in (b) and (c).}
\label{6_15_npoint_delta}
\end{figure*}

\begin{equation}
\phi_l(z)= \frac{1}{\Lambda \sqrt{\pi l!}} \left( \frac{z}{\Lambda} \right)^l
e^{-zz^*/(2\Lambda^2)},
\end{equation}
with complex coordinates $z=x+iy$. Namely, 
\begin{equation}
\Psi(z_1,z_2,...,z_N) = \sum_{J=1}^K C_J \Phi_J,
\label{exdwf}
\end{equation}
where $\Phi_J$ is a {\it normalized\/} permanent (see the Appendix); 
the index $J$ denotes any set of $N$ single-particle states
$\{\phi_{l_1}(z_1),\phi_{l_2}(z_2),...,\phi_{l_N}(z_N)\}$ (not necessarily 
distinct) with angular momenta $\{l_1,l_2,...,l_N\}$ such that 
\begin{equation}
l_1 \leq l_2 \leq ... \leq l_N,
\;\;\;\; {\text{and}} \;\;\;\; \sum_{k=1}^N l_k = L.
\label{sldc}
\end{equation}

The  dimension $K$ of the Hilbert space is controlled by the 
maximum allowed single-particle angular momentum $l_{\text{max}}$, such that 
$l_k \leq l_{\text{max}}$, $1 \leq k \leq N$. 
By varying $l_{\text{max}}$, we have checked that this procedure produces well 
converged numerical results.

We probe the intrinsic structure of the EXD eigenstates [Eq.\ (\ref{exdwf})]
for crystalline behavior using the conditional probability distribution 
$P({\bf r},{\bf r}_{0})$ defined (see, e.g., Ref.\ \cite{yl8}) by the following 
expression:
\begin{equation}
P({\bf r},{\bf r}_{0})=\langle \Psi|\sum_{i=1}^{N}\sum_{j \neq i}^{N} 
\delta({\bf r}_{i}-{\bf r})\delta({\bf r}_{j}-{\bf r}_{0})|\Psi \rangle.
\label{cpd_definition}
\end{equation}

Unlike the case of electrons in QDs \cite{yl4}, we found that, for rotating
bosons, the CPDs are not sufficient to decipher the intrinsic molecular 
configuration in the case of high fractional fillings $ \nu \geq 1/2$. In such
a case, one needs to calculate higher correlation functions. In this paper,
we use the $N$-point correlation function defined as the modulus square of the 
full many-body EXD wave function, i.e., 
\begin{equation}
P({\bf r};{\bf r}_1,{\bf r}_2,...,{\bf r}_{N-1})=
| \Psi({\bf r};{\bf r}_1,{\bf r}_2,...,{\bf r}_{N-1})|^2,
\label{npoint_def}
\end{equation}
where one fixes the positions of $N-1$ particles and inquiries about the 
(conditional) probability of finding the $N$th particle at any position 
${\bf r}$.

In the rest of this paper, we often find useful to contrast the CPDs and
$N$-point correlations to the single-particle densities $n({\bf r})$, which are 
circularly symmetric as a result of the total angular momentum being a good 
quantum number. The single-particle density is defined as
\begin{equation}
n({\bf r}) = 
\langle \Psi | \sum_{i=1}^N \delta({\bf r}_i-{\bf r}) | \Psi \rangle.
\label{spdensity}
\end{equation}

\section{Formation of rotating boson molecules: Numerical EXD results}
\label{results}
\subsection{The N=6 case} 
Since $H_{\text{LLL}}$ is rotationally invariant, i.e., $[H,L]=0$, its 
eigenstates 
$\Psi_L$ must also be eigenstates of angular momentum with eigenvalue $\hbar L$.
For a given rotational frequency $\Omega$, the eigenstate with lowest
energy is the ground state; we denote the corresponding angular momentum
as $L_{\text{gs}}$.
 
We proceed to describe our results for $N=6$ particles interacting via a
Coulomb repulsion by referring to  
Fig.\ \ref{6_m_vs_om}, where we plot $L_{\text{gs}}$ against the angular 
frequency $\Omega$. A main result from all our calculations is that 
$L_{\text{gs}}$ increases in characteristic (larger than unity) steps that take 
only a few integer values, i.e., for $N=6$ the variations of $L_{\text{gs}}$ are
in steps of 5 or 6. In keeping with
previous work on electrons \cite{yl1,yl2,yl3,yl4,yl5,yl6} at high $B$, 
and very recently on bosons in rotating traps \cite{yl11,yl12,yl13,mann,barb}, 
we explain these {\it magic-angular-momenta\/}  patterns (i.e., for $N=6$, 
$L_{\text{gs}} = L_0+5k$ or $L_{\text{gs}} 
= L_0+6k$, with $L_0=0$) as manifestation of an {\it intrinsic\/} point-group 
symmetry associated with the many-body wave function. This point-group symmetry 
emerges from the formation of {\it rotating boson molecules\/} (RBMs), 
i.e., from the localization of the bosons at the vertices of concentric regular 
polygonal rings; it dictates that the angular momentum of a purely rotational
state can only take values $L_{\text{gs}}=L_0+k_i n_i$, 
where $n_i$ is the number of localized particles on the $i$th polygonal ring 
\cite{note3}. Thus for $N=6$ bosons, the series $L_{\text{gs}}=5k$ is 
associated with an $(1,5)$ polygonal ring structure, while the series 
$L_{\text{gs}}=6k$ relates to an $(0,6)$ arrangement of particles. 
It is interesting to note that in classical calculations \cite{peeters_classical}
for $N=6$ particles in a harmonic 2D trap, the $(1,5)$ arrangement is found to
be the global energy minimum, while the $(0,6)$ structure is the lowest
metastable isomer. This fact is apparently reflected in the smaller
weight of the $L_{\text{gs}}=6k$ series compared to the $L_{\text{gs}}=5k$
series, and the gradual disappearance of the former with increasing $L$.   

\begin{figure}[t]
\centering
\includegraphics[width=7.5cm]{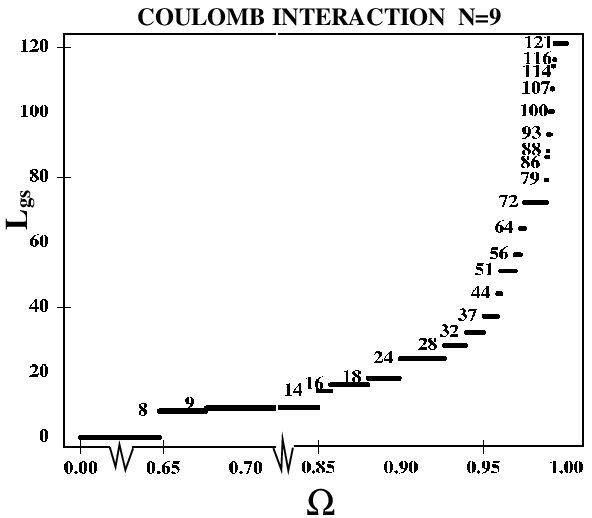}
\caption{Ground-state angular momenta, $L_{\text{gs}}$, for $N=9$ bosons in a 
rapidly rotating trap [described by the LLL Hamiltonian in Eq.\ (\ref{H_lll})], 
as a function of the rotational frequency $\Omega$ expressed in units of
$\omega_\bot$. The 
bosons interact via a Coulombic repulsion and the many-body Hilbert space is 
restricted to the LLL. The angular momentum associated with the first bosonic 
Laughlin state occurs at $L=72$, i.e., at $N(N-1)$. The value of 
$c=0.2 \hbar \omega_\bot \Lambda$ for this plot; 
the many-body wave functions do not depend on this choice.
Unlike the case of a $\delta$-interaction
(see Fig.\ \ref{9_dlt_vs_om}), the values of the ground-state angular momenta do
not terminate at $L=72$ (first Laughlin state), but they diverge as 
$\Omega/\omega_\bot \rightarrow 1$. Note the stepwise variation of the values 
of the ground-state angular momenta, indicating the presence of
an {\it intrinsic\/} point-group symmetry associated with the polygonal-ring
structure of RMBs (see text for details).}
\label{9_clmb_vs_om}
\end{figure}

\begin{figure}[t]
\centering
\includegraphics[width=7.5cm]{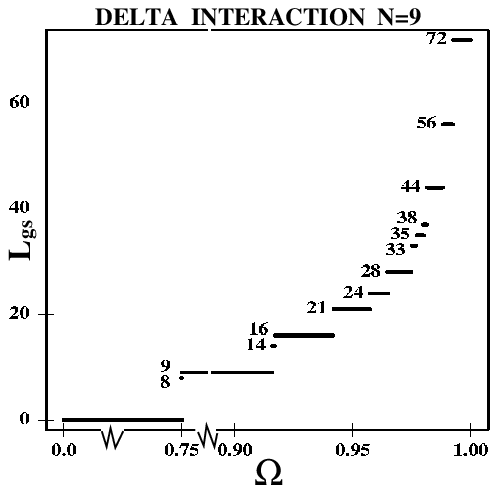}
\caption{Ground-state angular momenta, $L_{\text{gs}}$, for $N=9$ bosons in a 
rapidly rotating trap [described by the LLL Hamiltonian in Eq.\ (\ref{H_lll})], 
as a function of the rotational frequency $\Omega$ expressed in units of
$\omega_\bot$. The 
bosons interact via a {\it delta\/} repulsion and the many-body Hilbert space is 
restricted to the LLL. The angular momentum associated with the first bosonic 
Laughlin state occurs at $L=72$, i.e., at $N(N-1)$. The value of 
$g=2\pi \hbar \omega_\bot \Lambda^2/N$ for this plot; 
the many-body wave functions do not depend on this choice.
Unlike the case of a Coulomb 
interaction (see Fig.\ \ref{9_clmb_vs_om}), the values of the ground-state 
angular momenta do terminate at $L=72$ (first Laughlin state) when 
$\Omega/\omega_\bot = 1$. Note the stepwise variation of the values 
of the ground-state angular momenta, indicating the presence of
an {\it intrinsic\/} point-group symmetry associated with the polygonal-ring
structure of RMBs (see text for details).}
\label{9_dlt_vs_om}
\end{figure}

Magic values dominate also the ground state angular momenta of neutral bosons 
(delta repulsion) in rotating traps, as shown for $N=6$ bosons in Fig.\
\ref{6_m_vs_om_dlt}. Although the corresponding $\Omega$-ranges along the 
horizontal axis may be different compared to the Coulomb case, 
the appearance of only
the two series $5k$ and $6k$ is remarkable -- pointing to the formation of
RBMs with similar $(1,5)$ and $(0,6)$ structures in the case of a delta 
interaction as well (see also Refs.\ \cite{mann,barb}). An importance 
difference, however, is that for the delta interaction both series end 
at $\Omega/\omega_\bot=1$ with the value $L=N(N-1)=30$ (the bosonic Laughlin
value at $\nu=1/2$), while for the Coulomb interaction this $L$ value is reached
for $\Omega/\omega_\bot<1$ -- allowing for an infinite set of magic angular 
momenta [larger than $N(N-1)$] to develop as $\Omega/\omega_\bot \rightarrow 1$. 
Later, we will return back to this important difference between long range
and short range interactions.

\begin{figure*}
\centering
\includegraphics*[width=15cm]{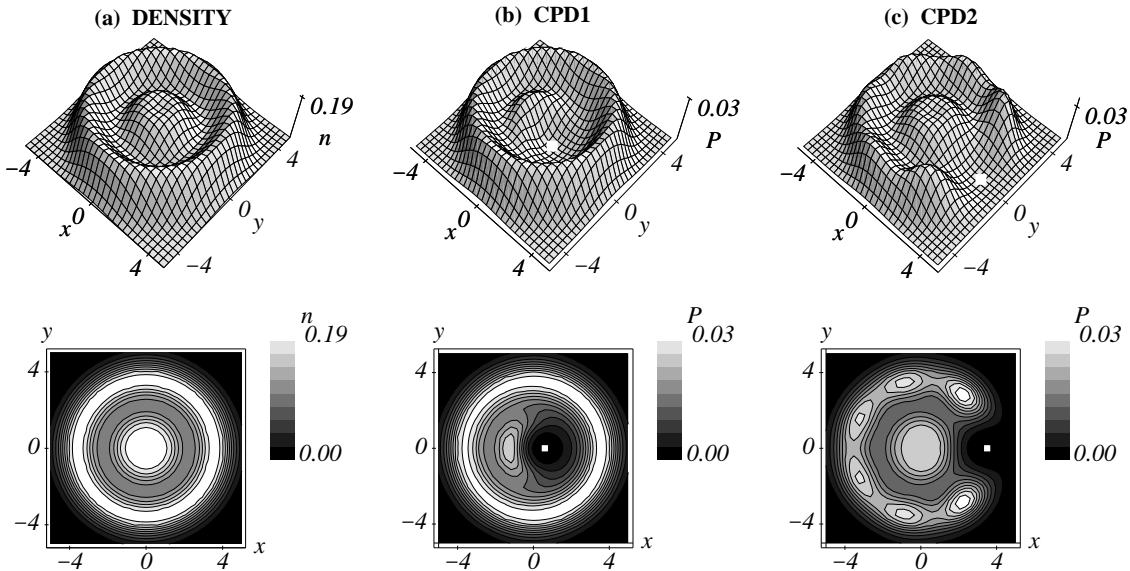}
\caption{Single-particle density and CPDs for the ground state of $N=9$ bosons 
interacting via a Coulomb repulsion and having a total angular momentum of 
$L_{\text{gs}}=93$. 
The RBM crystalline structure corresponds to a $(2,7)$ polygonal-ring 
arrangement (which is the most stable classical equilibrium structure). 
Top row: contour plots. Bottom row: 3D plots. 
By column we describe the individual panels in both rows as follows:   
(a) Single-particle density. (b) CPD1 with reference point (indicated by a white
dot) on the inner ring, revealing the localization of a second antipodal boson 
on the same inner ring. The outer ring exhibits a circular uniform symmetry that
reflects the independent rotation of the two rings. (c) CPD2 with reference 
point on the outer ring revealing the localization of 6 bosons relative to the 
reference boson. We note that now the inner ring exhibits a uniform circular 
symmetry, reflecting again the independent rotation of the two rings.
Lengths in units of $\Lambda$. The vertical scales are arbitrary, but
the same in (b) and (c).}
\label{inner_outer_9}
\end{figure*}

Beyond the analysis of the ground-state spectra as a function of $\Omega$,
the intrinsic crystalline point-group structure can be revealed by an
inspection of the CPDs [and to a much lesser extent by an inspection of
single-particle densities]. 
Because the EXD many-body wave function is an eigenstate of the
total angular momentum, the single-particle densities are circularly symmetric
and can only reveal the presence of concentric rings through oscillations
in the radial direction. The localization of bosons within the same ring
can only be revealed via the azimuthal variations of the anisotropic CPD 
[Eq. (\ref{cpd_definition})]. One of our findings is 
that for a given $N$ the crystalline features 
in the CPDs develop slowly as $L$ increases (or $\nu$ decreases). 

For $\nu < 1/2$, we find that the crystalline features are well developed for all
sizes studied by us. 
In Fig.\ \ref{6_coulomb_crystallization} and Fig.\ \ref{6_isomer}, we present
some concrete examples of CPDs from EXD calculations
associated with the ground-states of 
$N=6$ bosons in a rotating trap interacting via a repulsive Coulomb potential. 
In Fig.\ \ref{6_coulomb_crystallization}, we present the CPDs for $L_{\text{gs}}=
30$ (bosonic Laughlin for $\nu=1/2$), 60, and 120; these angular momenta are
associated with ground states at specific $\Omega$-ranges 
[see Fig.\ \ref{6_m_vs_om}].
All three of these angular momenta are divisible by both 5 and 6. However,
only the $L_{\text{gs}}=30$ CPD (Fig.\ \ref{6_coulomb_crystallization} 
top row) has a structure that is
intermediate between the $(1,5)$ and the $(0,6)$ polygonal-ring arrangements.
The two other CPDs, associated with the higher $L_{\text{gs}}=60$ and
$L_{\text{gs}}=120$ exhibit clearly only the $(1,5)$, illustrating our
statement above that the quantum-mechanical CPDs conform to the structure of
the most stable arrangement (i.e., the $(1,5)$ for $N=6$) of classical 
point-like charges as the fractional filling decreases. On the other hand,
in Fig.\ \ref{6_isomer}, we plot the CPD for $L_{\text{gs}}=36$, 
which is divisible only by 6; we see that the corresponding single-particle 
density and CPD are associated with a $(0,6)$ polygonal-ring arrangement.

However, for $\nu > 1/2$, the azimuthal variations may
not be visible in the CPDs, in spite of the characteristic step-like 
ground-state spectra [see Fig.\ \ref{6_m_vs_om} for $N=6$ bosons and 
Fig.\ \ref{9_clmb_vs_om} for $N=9$ bosons]. This paradox is resolved when one
considers higher-order correlations, and in particular $N$-point correlations
[see Eq.\ (\ref{npoint_def})]. In Fig.\ \ref{6_12_npoint_clmb} and Fig.\ 
\ref{6_12_npoint_delta}, we plot the $N$-point correlation functions for
$N=6$ bosons and $L_{\text{gs}}=12$ for both the Coulomb interaction and 
$\delta$-repulsion, 
respectively, and we compare them against the corresponding CPDs. The value
of 12 is divisible by 6, and one expects this state to be associated with
a $(0,6)$ molecular configuration. It is apparent that the CPDs
fail to portray such sixfold azimuthal pattern. The $(0,6)$ pattern, however,
is clear in the $N$-point correlations, where we fixed five points (white dots)
at the positions ${\bf r}_k=r_0 \exp(i k \pi/3)$, $k=1,2,...,5$ ($r_0$ being the 
radius of the maximum ring in the single-particle density) and we looked for the
probability of finding the sixth boson at any other position ${\bf r}$. 
The figures show that the maximum probability happens for 
${\bf r}_6=r_0 \exp(2\pi i)$ (black dot) which completes the $(0,6)$ regular 
polygon.

\begin{figure*}
\centering
\includegraphics*[width=15cm]{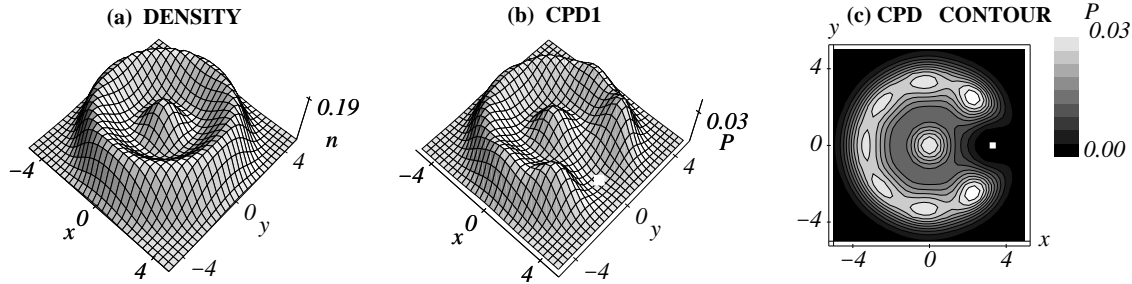}
\caption{A ground state with a $(1,8)$ RBM arrangement for $N=9$ bosons 
interacting via a Coulomb repuslion and having angular momentum 
$L_{\text{gs}}=88$. In the classical calculation \cite{peeters_classical}, 
this arrangement is isomeric to the most stable $(2,7)$ one exhibited by
the CPDs in Fig.\ \ref{inner_outer_9}. The individual panels are as follows: 
(a) $3D$ plot of the single-particle density. (b) $3D$ plot of the CPD. 
(c) Contour plot of the CPD. In the CPD plots, the white dots indicate the 
reference point ${\bf r}_0$. Lengths in units of $\Lambda$. The vertical scales 
are arbitrary, but the same in (b) and (c).}
\label{9_isomer}
\end{figure*}

\begin{figure*}
\centering
\includegraphics*[width=15cm]{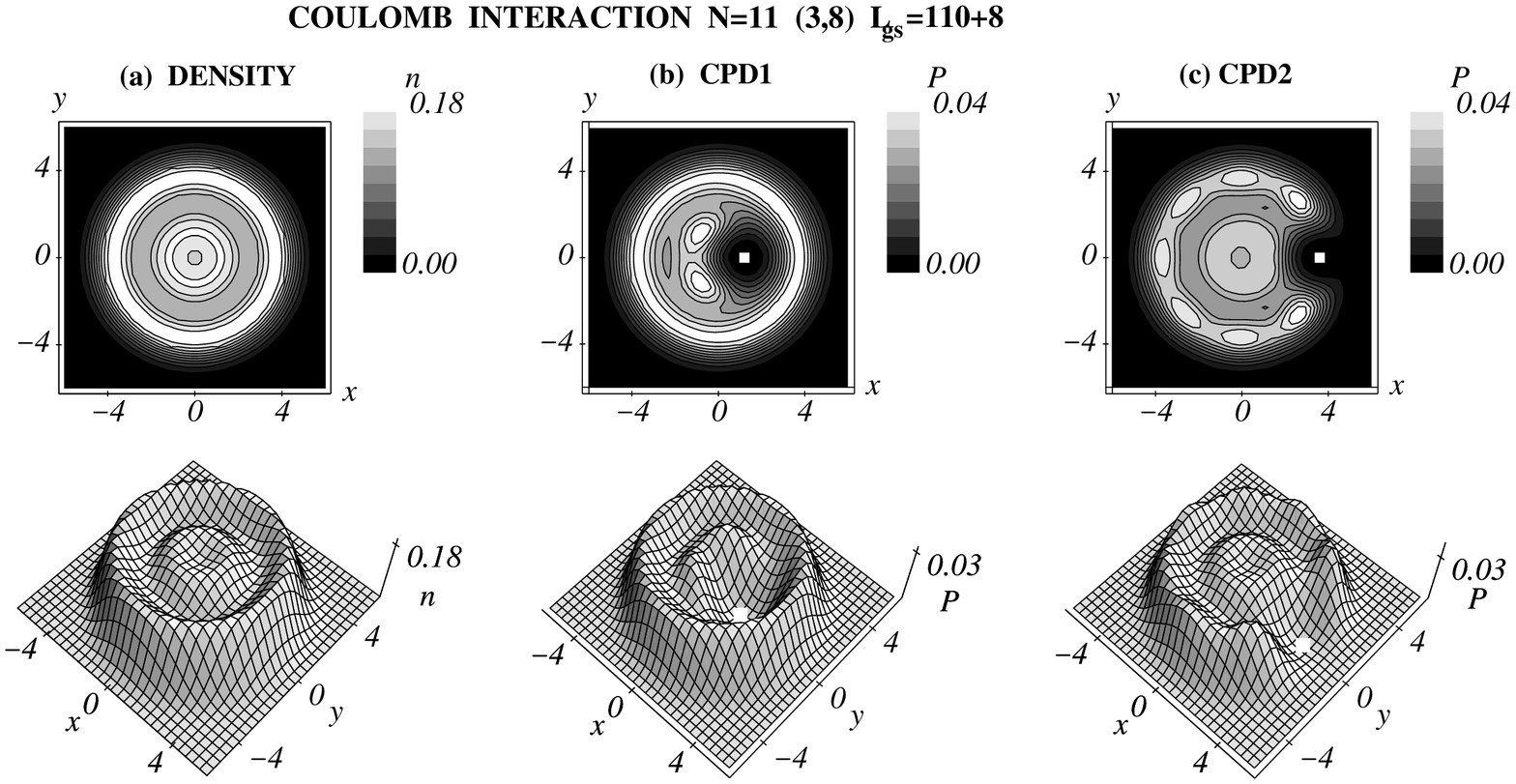}
\caption{Single-particle density and CPDs for the ground state of $N=11$ bosons 
interacting via a Coulomb repulsion and having a total angular momentum of 
$L_{\text{gs}}=118$. The RBM crystalline structure corresponds to a $(3,8)$ 
polygonal-ring arrangement (which is the most stable classical equilibrium 
structure \cite{peeters_classical}). 
Top row: contour plots. Bottom row: 3D plots. 
By column we describe the individual panels in both rows as follows:   
(a) Single-particle density. (b) CPD with reference point (indicated by a white 
dot) on the inner ring, revealing the localization of two bosons on the same 
inner ring relative to the reference boson. The outer ring exhibits a circular 
uniform symmetry that reflects the independent rotation of the two rings. 
(c) CPD with reference point on the outer ring revealing the localization of 7 
bosons relative to the reference boson. We note that now the inner ring exhibits
a uniform circular symmetry, reflecting again the independent rotation of the 
two rings (see text for details). Lengths in units of $\Lambda$. The 
vertical scales are arbitrary, but the same in (b) and (c).}
\label{inner_outer_11}
\end{figure*}

A second example is given in Fig.\ \ref{6_15_npoint_clmb} and Fig.\
\ref{6_15_npoint_delta} for the case of $N=6$ bosons and $L_{\text{gs}}=15$,and 
for both the Coulomb interaction and the $\delta$-repulsion, respectively. In 
this second case, one expects a $(1,5)$ molecular pattern, which however is not
visible in the CPDs. To reveal this $(1,5)$ pattern, one needs to plot the
$N$-point correlations (middle and right panels). One now has two choices for
choosing the positions of the first five particles (white dots), i.e., one 
choice places one white dot at the center and the other choice places all five
white dots on the vertices of a regular pentagon. For both choices, as shown by 
the contour lines in the figures, the position of maximum probability for the 
sixth boson coincides with the point that completes the $(1,5)$ configuration
[see the black dots at ${\bf r}_6=r_0 \exp(2\pi i)$ in the middle panels and 
at the center in the right panels].

\begin{figure*}
\centering
\includegraphics*[width=15cm]{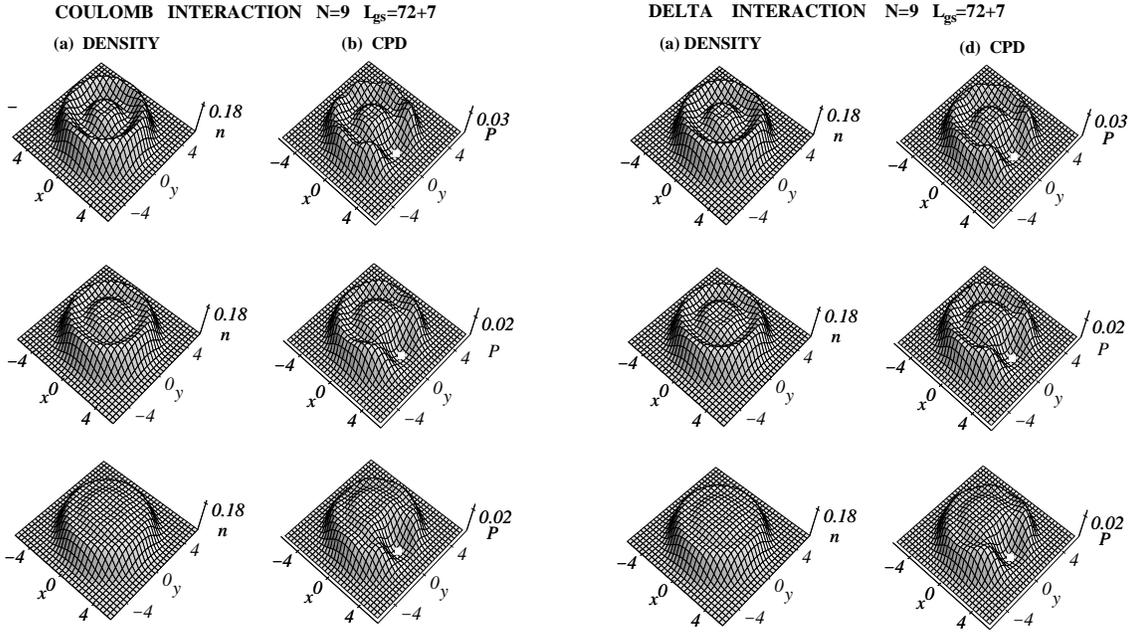}
\caption{Comparison of single-particle densities and CPDs for three degenerate 
states (right half of figure) of $N=9$ bosons interacting via a repulsive 
$\delta$-interaction and having 
$L_{\text{gs}}=79$ against the lowest three (and non-degenerate) eigenstates 
(left half of figure) of the Coulomb repulsion for the same 
angular momentum. The numbers on the left hand side of the rows 
indicate the order of the eigenstates in the case of the Coulomb interaction,
with 1 (top row) indicating the ground state, while 2 and 3 indicate excited
(rovibrational \cite{yl7}) states. The top-row plots reveal a $(2,7)$ 
crystalline $(2,7)$ structure for both types (long-range and short-range) 
interactions. Lengths in units of $\Lambda$. The vertical scales are 
arbitrary, but the same for the single-particle densities [(a) and (c)] and for
the CPDs [(b) and (d)].}
\label{clmb_delta_79}
\end{figure*}

Note that for both cases, the differences in the CPDs and $N$-point 
correlation functions between the Coulomb and the $\delta$-repulsion are
rather minimal. We will return to this result again in Section III.D. 

\subsection{The $N=9$ case with Coulomb repulsion}

The results for $N=6$ bosons portray many of the important features for all the 
$N$'s that we have examined, keeping in mind that for $N=6$ the corresponding 
polygonal-ring structures involve only a single {\it nontrivial\/} ring, i.e., 
they are of the form $(1,5)$ or $(0,6)$. 
From the case of confined electrons \cite{yl4,yl5}, it 
is known that the first nontrivial two-ring structure appears for $N=9$. Our 
results presented here confirm that this is also the case for $N=9$ bosons.

In Fig.\ \ref{9_clmb_vs_om}, we display the LLL ground-state angular momenta
obtained from EXD calculations
for $N=9$ bosons as a funtion of $\Omega/\omega_0$ and for the case of
a Coulombic repulsion. The remarkable feature again is the variation of 
$L_{\text{gs}}$ in well defined steps, with the steps taking only three 
specific values 7, 8, and 9 (with the step value of 9 appearing only in two 
instances). The coresponding
magic-angular-momenta values correspond to the series $L_{\text{gs}}=2k_1+7k_2$
(${\cal S}_7$), $L_{\text{gs}}=8k$ (${\cal S}_8$), and $L_{\text{gs}}=9k$ 
(${\cal S}_9$), associated with the polygonal-ring configurations 
$(2,7)$, $(1,8)$, and $(0,9)$, respectively. Note that, as $\Omega/\omega_0
\rightarrow 1$, only the ${\cal S}_7$ series survives, a behavior that is again
related to the fact that the $(2,7)$ structures is the most stable isomer
according to classical elecrostatic calculations \cite{peeters_classical},
with the $(1,8)$ structure being the lowest metastable isomer.   

In Fig.\ \ref{inner_outer_9}, we present the single-particle density and CPD for
$N=9$ bosons and $L_{\text{gs}}=93$ (with $k_1=1$ and $k_2=13$). 
Naturally, the single-particle density
is circularly symmetric, but it clearly protrays two concentric rings along
the radial direction. Furthermore, the$(2,7)$ intrinsic azimuthal crystalline 
pattern is revealed in the two displayed CPDs, one with the observation point 
being located on the inner ring (middle panels) and the other on the outer ring 
(right panels). It is remarkable that positioning
the reference point ${\bf r}_0$ on one ring reveals only the intrinsic
crystalline correlations of the same ring, while the other ring appears
uniform, and {\it vice versa\/}. This behavior is similar to that of multi-ring
rotating electron molecules \cite{yl4,yl5}, and it suggests that the rings of the
RBMs rotate {\it independently\/} from each other. It naturally leads to a 
physical picture of the RBMs as highly {\it nonrigid\/} rotors in analogy with 
the analysis \cite{yl4,yl5} of the rotating electron molecules in high $B$ 
(see in particular sections V and VI in Ref.\ \cite{yl5}).   

We close this subsection by showing a case associated the second isomer.
Fig.\ \ref{9_isomer} displays the single-particle density and CPDs for $N=9$ 
bosons and $L_{\text{gs}}=88$ (belonging to the ${\cal S}_8$ series with $k=11$).
Both of these quantities confirm that the ground state has a $(1,8)$ 
intrinsic crystalline pattern, in agreement with the analysis above regarding
the ground-state angular-momentum spectrum in Fig.\ \ref{9_clmb_vs_om}.

\begin{figure*}
\centering
\includegraphics*[width=15cm]{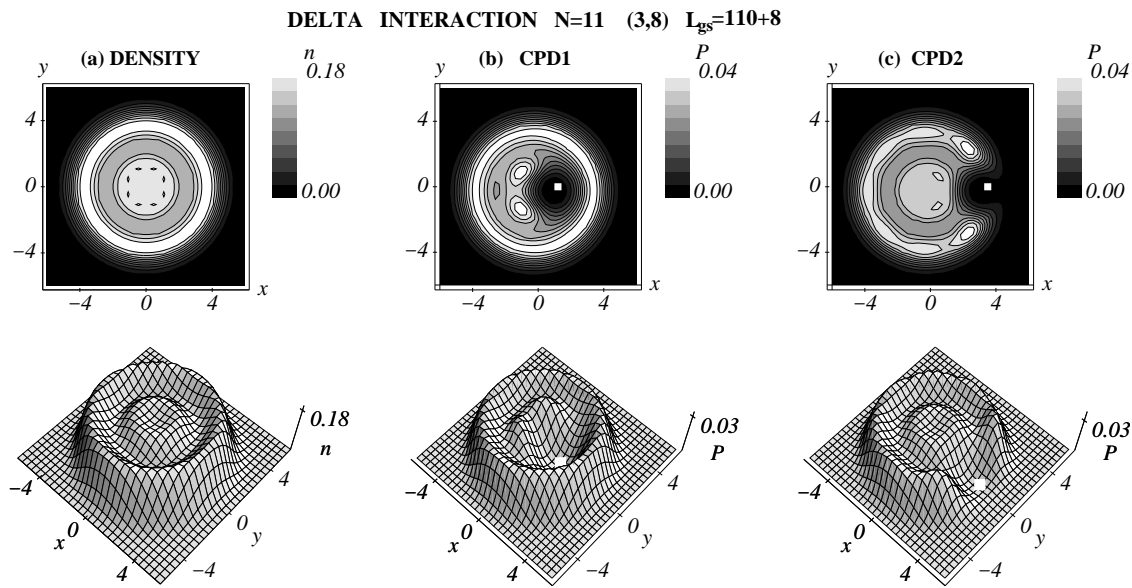}
\caption{Single-particle density and CPDs for one of the degenerate ground state
of $N=11$ bosons interacting via a {\it delta} repulsion and having a total 
angular momentum of $L_{\text{gs}}=118$. The RBM crystalline structure 
corresponds to a $(3,8)$ polygonal-ring arrangement (which is the most stable 
classical equilibrium structure \cite{peeters_classical}). 
Top row: contour plots. Bottom row: 3D plots. 
By column we describe the individual panels in both rows as follows:   
(a) Single-particle density. (b) CPD with reference point 
(indicated by a white dot) on the 
inner ring, revealing the localization of two bosons on the same inner ring
relative to the reference boson. The outer ring exhibits a circular uniform 
symmetry that reflects the independent rotation of the two rings. (c) CPD with 
reference point on the outer ring revealing the localization of 7 bosons 
relative to the reference boson. We note that now the inner ring exhibits a 
uniform circular symmetry, reflecting again the independent rotation of the two 
rings (see text for details). Lengths in units of $\Lambda$. The 
vertical scales are arbitrary, but the same in (b) and (c).}
\label{inner_outer_dlt_11}
\end{figure*}

\subsection{The N=11 case with Coulomb repulsion}

Figure \ref{inner_outer_11} provides another example of the independent
rotation of the rings that form the RBMs. In particular, Fig.\ 
\ref{inner_outer_11} displays the single-particle density and the two CPDS 
(for the inner and outer ring) for $L_{\text{gs}}=118$ and $N=11$ bosons 
interacting via a Coulomb repulsive potential. The intrinsic crystalline 
structure has a $(3,8)$ pattern, in agreement with the decomposition of the 
total angular momentum as $118=3 \times 2 + 8 \times 14$ and the fact that the 
$(3,8)$ isomer is the most stable one according to classical electrostatic 
calculations
\cite{peeters_classical}.

\subsection{Short-range vs. long-range interactions}

In the present investigation, we also also studied the influence of the range of
the interaction on the properties of the many-body wavefunctions associated
with a finite number of rapidly rotating bosons. In particular, using both a 
Coulombic and a delta interaction, we have calculated and compared 
single-particle densities 
and CPDs for several sizes and fractional fillings $\nu$. Our result is that
both types of interactions yield similar CPDs, i.e., that the degree of 
crystalline correlations does not depend strongly on the range of the 
interaction, as long as one refers to the same size $N$ and fractional filling
$\nu$. 

For example, the well developed crystallinity for $\nu < 1/2$ familiar from the
Coulombic case (see earlier sections) maintains also for the case of a
$\delta$-interaction, as Fig.\ \ref{clmb_delta_79} ($N=9$) and Fig.\ 
\ref{inner_outer_dlt_11} ($N=11$) illustrate. 
In particular, Fig.\ \ref{clmb_delta_79} compares for $N=9$ and $L=93$ the 
single-particle densities and CPDs associated with the three lowest states 
in the case of the Coulomb interaction (left half of the figure) against
the corresponding quantities for three states out of the set of the now 
multiply degenerate \cite{vief} ground states in the case of the 
$\delta$-interaction. Of course, this multiple degeneracy is unphysical, and
in experimental applications one should use a short-range potential with a 
finite interaction length that is intermediate between the Coulomb and the 
$\delta$-potential, like the dipole-dipole interaction \cite{youl}. For our
purposes here, it is sufficient to view the $\delta$-interaction as a 
representative one of all short-range potentials, and thus to conclude that
RBMs with similar patterns and properties do form, 
and that the degree of crystallinity is strengthened as $\nu$ decreases,
for both long-range and short-range interactions.
We stress that the similarities involve both the 
crystalline arrangements [i.e., the $(2,7)$ polygonal-ring pattern] and the
property that the rings rotate independently of each other, as can be
seen by a direct comparison of the left and right halves of Fig.\ 
\ref{clmb_delta_79}. We note that certain similarities (regarding the formation
of RBMs in the LLL) between the long-range (Coulomb) and short-range (Gaussian
interaction) cases have also been reported \cite{mann} most recently in the
context of EXD/LLL studies of one-ring RBMs (with $N \leq 7$ and $\nu < 1/2$).
For similarities in the case $\nu \geq 1/2$, see Section III.A.

Finally Fig.\ \ref{inner_outer_dlt_11} displays the CPD 
associated with one of the degenerate 
ground states for $L=118$ and $N=11$ bosons interacting via a $\delta$-potential.
In keeping with our conclusions in the previous paragraph, we point out
the clear formation of an RBM with a $(3,8)$ crystalline pattern; furthermore,
this RBM also exhibits the property that the two rings rotate independently of 
each other.

\section{Conclusions}
\label{conclusion}
We presented results for the exact diagonalization of a small number $ N \leq 11$
of rapidly rotating bosons interacting via a repulsive $\delta$-function or 
Coulomb interaction. We focused on the {\em weakly} interacting limit where the 
assumption of confinement to the lowest Landau level is valid and studied 
in particular particle numbers which in the classical limit lead to polygonal
crystalline configurations with at least two concentric polygonal rings. For all 
fractional fillings, we observed formation of {\em rotating boson molecules},
i.e., the appearance of crystalline correlations in 
close similarity to those found for confined two dimensional electron 
systems in high $B$ \cite{yl1,yl2,yl3,yl4,yl5,yl6}, as well as for trapped  
{\em strongly} repelling bosons in the context of a recently developed
trial wave function \cite{yl11,yl13}. 

Our main result is that for small filling 
fractions, i.e., $\nu < 1/2$, the molecular character is well developed and
explicit in the CPDs; for larger fractions $\nu \geq 1/2$, the molecular
character is clearly revealed through inspection of higher-order correlation 
functions. The bosons organize into concentric polygonal rings which rotate 
{\em independently} of each other, as was earlier found \cite{yl4,yl5} for the 
case of rotating electron molecules in high $B$.  
The molecular structure is reflected in the ground-state spectra, i.e.,
as a function of the rotational frequency of the trap, 
the ground-state angular momenta vary in characteristic steps that coincide 
with the number of localized bosons on each polygonal ring \cite{note57}.
Comparison of results for $\delta$-function and Coulomb interactions also 
indicates that the emergence of the RBM is more dependent on the fractional 
filling $\nu$ than on the range of the inter-particle interaction.  

\acknowledgments

This research is supported by the U.S. D.O.E. (Grant No. FG05-86ER45234)
and the NSF (Grant No. DMR-0205328). Calculations were performed at the
Center for Computational Materials Science at Georgia Tech and at the
National Energy Research Scientific Computing Center (NERSC). 

\appendix
\section{Mathematical formalism and methods}

\subsection{formalism}
\label{norm}

Let 
\begin{equation} \Phi_{A}=\sqrt\frac{n_{1}!n_{2}!\hdots n_{M}!}{N!} 
\sum_{P}u_{P_{\alpha 1}}(1)u_{P_{\alpha 2}}(2)\hdots u_{P_{\alpha N}}(N),
\label{wavefunction}
\end{equation} 
represent one of the properly normalized permanents in Eq.\ (\ref{exdwf}) that 
form the basis of the many-body Hilbert space for the bosonic problem defined by
the Hamiltonian in Eq.\ (\ref{H_lll}). Such a permanent is constructed with 
$M (\leq N)$ orthonormal single particle orbitals $u_{\alpha_i}(i)$ (LLL 
states in our case). The orbitals are indexed by a single integer 
$\alpha_i$ and $(i)$ stands for all coordinates of particle $i$. While
$i=1,2,\hdots,N$, the indices $\alpha_i$ are not necessarily distinct.  
The summation $\sum_{P}$ runs over all {\em distinct} permutations of $N$ 
particles among the $M$ orbitals with occupations (multiplicities) 
$\{n_1,n_2,\hdots,n_M\}$. 

It is straightforward to show that the permanent defined in Eq.\ 
(\ref{wavefunction}) is normalized to unity. Indeed its norm is given  by the 
following expression:
\begin{eqnarray}
&&(\Phi_A^*,\Phi_A)=\int\prod_{i=1,N} d{\bf r}_i
\frac{n_{1}!n_{2}!\hdots n_{M}!}{N!} \nonumber \\
&&\times \sum_{P} \sum_{Q} u_{P_{\alpha_{1}}}^{*}(1) u_{Q_{\alpha_{1}}}(1) \dots 
u_{P_{\alpha_{N}}}^{*}(N) u_{Q_{\alpha_{N}}}(N),\qquad \nonumber\\
\label{phioverlap}
\end{eqnarray}
which upon evaluating the integrals yields
\begin{eqnarray}
(\Phi_{A}^{*},\Phi_{A})&=&
\frac{n_{1}!n_{2}!\hdots n_{M}!}{N!}\sum_{P} \sum_{Q}
\delta_{{P_{\alpha_{1}}},{Q_{\alpha_{1}}}} \dots 
\delta_{P_{\alpha_{N}}, Q_{\alpha_{N}}}
\nonumber \\   
&=& \frac{n_{1}!n_{2}!\hdots n_{M}!}{N!}\sum_{P} (1)  \nonumber \\ 
&=& \frac{n_{1}!n_{2}!\hdots n_{M}!}{N!} 
\frac{N!}{n_{1}!n_{2}!\hdots n_{M}!}\nonumber =1 
\label{norm_evaluation}
\end{eqnarray}
where in the last line the summation equals the number of distinct ways of 
arranging $N$ particles among $M$ orbitals with the $j$th orbital having
multiplicity $n_{j}$.

\subsection{Interaction Matrix Elements}

We make use of $\Phi_A$, i.e., Eq.\ (\ref{wavefunction}) above, which represents
a properly symmetrized and normalized wavefunction to calculate the matrix 
elements for the many-body interaction term which we denote here as 
\begin{equation}
V_{\text{\text{I}}} = \sum_{i < j} v_{ij}. 
\end{equation}
At the outset we note that $v_{ij}=v({\bf r}_i-{\bf r}_j)$ is symmetric with
respect to the exchange of particles $i$ and $j$ and  due to the symmetry of 
bosonic wavefunctions, this property is carried over to the expression: 
\begin{equation}
(\Phi_{A}^{*}
({\bf r}_1,{\bf r}_2,\hdots,{\bf r}_N),v_{ij} 
\Phi_{B}({\bf r}_1,{\bf r}_2,\hdots,{\bf r}_N))
\label{interaction_0}
\end{equation}
In other words Eq.\ (\ref{interaction_0}) 
doesn't change value under exchange of all the coordinates of particles $i$ 
and $j$ with any other pair.
So 
\begin{equation}
\langle V_{\text{I}} \rangle_{AB}=
\sum_{i < j} (\Phi_{A}^{*},v_{ij} \Phi_{B})=  
\frac{N(N-1)}{2}(\Phi_{A}^{*},v_{12} \Phi_{B}) 
\end{equation}
and the matrix element for the many-body Hamiltonian due to the interaction
is reduced to
\begin{equation}
\langle V_{\text{I}} \rangle_{AB}=
\frac{N(N-1)}{2}(\Phi_{A}^{*},v_{12} \Phi_{B})
\label{interaction_1}
\end{equation}

From our calculation of the norm in section~\ref{norm} above, it should be 
obvious that matrix element defined by Eq.\ (\ref{interaction_1}) can be 
non-zero if $\Phi_A$ and $\Phi_B$ are the same or differ in the occupations 
of two orbitals from each one; the case when $\Phi_A$ and $\Phi_B$ differ in the
occupations of one orbital from each permanent does not occur in the LLL.
For all other cases the matrix elements in Eq.\ (\ref{interaction_1}) are zero. 
Because of the normalization factors introduced by the occupation numbers 
$n_{j}$, the evaluation of Eq.\ (\ref{interaction_1}) breaks down into five 
cases, one for the diagonal elements and four for the off-diagonal elements. We 
shall take them in turn. 

\subsection{Diagonal Elements}
Substituting the expression for the normalized wavefunction 
Eq.\ (\ref{wavefunction}) into (\ref{interaction_1}) above, we obtain
\begin{widetext}
\begin{eqnarray}
\langle V_{\text{I}} \rangle_{AA} &=&
\frac{N(N-1)}{2}\int\prod_{i=1,N}d{\bf r}_i
\frac{n_{1}!\hdots n_{M}!}{N!}\sum_{P,Q} 
u_{P_{\alpha_{1}}}^{*}(1)u_{P_{\alpha_{2}}}^{*}(2)v_{12} 
u_{Q_{\alpha_{1}}}(1)u_{Q_{\alpha_{2}}}(2) 
\prod_{j=3,N}u_{P_{\alpha_{j}}}^{*}(j) 
u_{Q_{\alpha_{j}}}(j)\qquad \nonumber \\
&=& \frac{N(N-1)}{2} \int d{\bf r}_1 d{\bf r}_2
\frac{n_{1}!n_{2}! \ldots n_{M}!}{N!} \sum_{P,Q} 
u_{P_{\alpha_{1}}}^{*}(1)u_{P_{\alpha_{2}}}^{*}(2)v_{12} 
u_{Q_{\alpha_{1}}}(1)u_{Q_{\alpha_{2}}}(2) 
\prod_{i=3,N}\delta_{P_{\alpha_{i}}, Q_{\alpha_{i}}} 
\label{interaction_2}
\end{eqnarray}
\end{widetext}
In Eq.\ (\ref{interaction_2}) only two terms in the sum $\sum_Q$ can be nonzero,
i.e., the terms satisfying the two conditions below 
\begin{eqnarray}
&&P_{\alpha_{1}} = Q_{\alpha_{1}} \;\text{and}\; P_{\alpha_{2}}=Q_{\alpha_{2}},
\;\;\text{or} \nonumber \\
&&P_{\alpha_{1}} = Q_{\alpha_{2}} \; \text{and} \; P_{\alpha_{2}}=Q_{\alpha_{1}};
\nonumber\\
&&\hspace{-2.2cm} \text{and for both of the above cases} \nonumber \\ 
&&P_{\alpha_j} = Q_{\alpha_j},\;\;\; j=3,\ldots,N.
\label{condi}
\end{eqnarray}

Thus  Eq.\ (\ref{interaction_2}) is rewritten as
\begin{eqnarray}
&&\langle V_{\text{I}} \rangle_{AA} = 
\frac{n_{1}!\ldots n_\alpha! n_\beta!\hdots n_{M}!}{2(N-2)!} \nonumber \\
&&\times
\sum_{\alpha\beta} \biglb( \langle \alpha \beta |v_{12}| \alpha \beta \rangle + 
\langle \alpha \beta |v_{12}| \beta \alpha  \rangle \bigrb)
\sum_{ P\{N-2\} } (1), \nonumber \\
\label{inter3}
\end{eqnarray}
where we have simplified the notation of the indices, i.e., we set
$P_{\alpha_{1}}=\alpha$ and $P_{\alpha_{2}}=\beta$, and we used 
the following shorthand definition for the matrix elements of the two-body
interaction
\begin{equation}
\langle \alpha \beta|v_{12}|\gamma \Delta \rangle=
\int d{\bf r}_{1}d{\bf r}_{2} 
u_{\alpha}^{*}(1)u_{\beta}^{*}(2)v_{12} u_{\gamma}(1)u_{\Delta}(2).
\end{equation}

In Eq.\ (\ref{inter3}), we have explicitly noted the fact that the sum
$\sum_P$ refers now to $N-2$ particles. There are two case to be considered.
When $\alpha=\beta$, one has
\begin{equation}
\sum_{ P\{ N-2 \} } (1)= 
\frac{(N-2)!}{n_1! \ldots (n_\alpha-2)!n_\beta! \ldots n_M!}.
\end{equation}
When $\alpha \neq \beta$, one has
\begin{equation}
\sum_{ P\{ N-2 \} } (1)= 
\frac{(N-2)!}{n_1! \ldots (n_\alpha-1)!(n_\beta-1)! \ldots n_M!}.
\end{equation}

With the above results,  Eq.\ (\ref{inter3}) finally yields for the diagonal 
elements of the interaction part of the Hamiltonian

\begin{eqnarray}
\langle V_{\text{I}} \rangle_{AA}  &=& 
\frac{n_{\alpha}(n_{\alpha}-1)}{2} \sum_{\alpha=1}^M
\langle \alpha \alpha|v_{12}|\alpha \alpha \rangle 
\nonumber \\
&+& \frac{n_{\alpha}n_{\beta}}{2} \sum_{\alpha \neq \beta=1}^M
\biglb(\langle \alpha \beta|v_{12}|\alpha \beta \rangle  
+ \langle \alpha \beta|v_{12}|\beta \alpha \rangle \bigrb) \nonumber \\ 
\label{interaction_diagonal}
\end{eqnarray}
 
\subsection{Off-Diagonal Elements}

Here, we extend the formalism developed in the previous sections to be able to 
treat matrix elements between two different basis permanents.
Let $\Phi_{\alpha\beta}^{\gamma\Delta}$ and 
$\Phi_{\alpha-1,\beta-1}^{\gamma+1,\Delta+1}$ denote permaments having
the occupation numbers
\begin{equation}
\{n_{1},n_{2},\ldots,n_{\alpha},n_{\beta},n_{\gamma},n_{\Delta},\ldots,n_{M}\}
\label{phi_occup1}
\end{equation}
and  
\begin{equation}
\{n_{1},n_{2},\ldots,n_{\alpha}-1,n_{\beta}-1,n_{\gamma}+1,
n_{\Delta}+1,\ldots,n_{M}\},
\label{phi_occup2}
\end{equation}
respectively. Then we need to calculate the matrix element
\begin{equation}
(\Phi_{\alpha-1,\beta-1}^{\gamma+1,\Delta+1},V_{\text{I}} 
\Phi_{\alpha\beta}^{\gamma\Delta}).
\label{off_diagonal_1}
\end{equation}


$\Phi_{\alpha-1,\beta-1}^{\gamma+1,\Delta+1}$ can be interpreted as the permanent
created from $\Phi_{\alpha\beta}^{\gamma\Delta}$ when two particles are taken 
from the single particle states labelled by $\alpha$ and $\beta$ and placed into
the states $\gamma$ and $\Delta$. Four cases arise; they are 
\begin{eqnarray}
\gamma \; \neq \; \Delta; \;\;    \alpha \; \neq \;\beta \nonumber \\ 
\gamma \; \neq \; \Delta; \;\; \alpha \;= \;\beta \nonumber \\
\gamma \; = \; \Delta; \;\; \alpha \; \neq \;\beta \nonumber \\ 
\gamma \;= \; \Delta; \;\;  \alpha \;= \;\beta 
\end{eqnarray}

\begin{widetext}
\noindent
{$\bullet$ Case I:\; $ \gamma \; \neq \; \Delta; \; \alpha \;\neq \;\beta $}. \\
We begin from Eq.\ (\ref{interaction_1}) and make adjustments for 
the difference in the occupation numbers to obtain:
\begin{eqnarray}
&&\hspace{-0.5cm} (\Phi_{\alpha-1,\beta-1}^{*\gamma+1, \Delta+1},V_{\text{I}} 
\Phi_{\alpha\beta}^{\gamma\Delta}) = \nonumber \\
&&\hspace{-0.5cm} 
\frac{N(N-1)}{2} \sqrt\frac{n_{1}!\ldots(n_{\alpha}-1)!(n_{\beta}-1)!
(n_{\gamma}+1)!(n_{\Delta}+1)! \ldots n_{M}!}{N!}
\sqrt\frac{n_{1}!\ldots n_{\alpha}!n_{\beta}!n_{\gamma}!
n_{\Delta}! \ldots n_{M}!}{N!} \nonumber \\
&& \hspace{-0.5cm} 
\times \biglb( \langle \gamma \Delta |v_{12}|\alpha \beta \rangle + 
\langle \gamma \Delta |v_{12}|\beta \alpha \rangle
+\langle \Delta \gamma|v_{12}|\alpha \beta \rangle + 
\langle \Delta \gamma |v_{12}|\beta \alpha \rangle \bigrb) 
\frac{(N-2)!}{n_1! \ldots (n_{\alpha}-1)!(n_{\beta}-1)!n_{\gamma}!n_{\Delta}!
\ldots n_M!}
\nonumber \\
\end{eqnarray}
or finally
\begin{eqnarray}
&&(\Phi_{\alpha-1,\beta-1}^{*\gamma+1, \Delta+1},V_{\text{I}}
\Phi_{\alpha\beta}^{\gamma\Delta}) = \nonumber \\
&& \sqrt{n_{\alpha}n_{\beta}(n_{\gamma}+1)(n_{\Delta}+1)}
\biglb( \langle \gamma \Delta |v_{12}|\alpha \beta \rangle + 
\langle \gamma \Delta |v_{12}|\beta \alpha \rangle \bigrb)
\label{case1}
\end{eqnarray}

With similar considerations we obtain the results for the other cases.

\noindent
{$\bullet$ Case II: $\gamma \; \neq \; \Delta; \; \alpha \;= \;\beta$.}
\begin{eqnarray}
&&(\Phi_{\alpha-2}^{*\gamma+1, \Delta+1},V_{\text{I}} 
\Phi_{\alpha}^{\gamma\Delta}) = \nonumber \\
&&\frac{N(N-1)}{2} \sqrt\frac{n_{1}!\ldots(n_{\alpha}-2)!
(n_{\gamma}+1)!(n_{\Delta}+1)! \ldots n_{M}!}{N!}
\sqrt\frac{n_{1}!\ldots n_{\alpha}!n_{\gamma}!
n_{\Delta}! \ldots n_{M}!}{N!} \nonumber \\
&& \times \biglb( \langle \gamma \Delta |v_{12}|\alpha \alpha \rangle + 
\langle \Delta \gamma |v_{12}|\alpha \alpha \rangle \bigrb)
\frac{(N-2)!}{n_1! \ldots (n_{\alpha}-2)!n_{\gamma}!n_{\Delta}!
\ldots n_M!} 
\nonumber \\
\end{eqnarray}
\end{widetext}
or finally
\begin{eqnarray}
&&(\Phi_{\alpha-2}^{*\gamma+1, \Delta+1},V_{\text{I}}
\Phi_{\alpha}^{\gamma\Delta}) = \nonumber \\
&&\sqrt{n_{\alpha}(n_{\alpha}-1)(n_{\gamma}+1)(n_{\Delta}+1)}
\langle \gamma \Delta |v_{12}|\alpha \alpha \rangle \nonumber\\
\label{case2}
\end{eqnarray} 

\noindent
{$\bullet$ Case III: $\gamma \; = \; \Delta; \; \alpha \; \neq \;\beta$.}
\begin{eqnarray}
&&(\Phi_{\alpha-1,\beta-1}^{*\gamma+2},V_{\text{I}}
\Phi_{\alpha\beta}^{\gamma}) = \nonumber \\
&&\sqrt{n_{\alpha}n_{\beta}(n_{\gamma}+1)(n_{\gamma}+2)}
\langle \gamma \gamma|v_{12}|\alpha \beta \rangle \nonumber\\
\label{case3}
\end{eqnarray} 

\noindent
{$\bullet$ Case IV: $\gamma \; = \; \Delta; \; \alpha \;= \;\beta$.}
\begin{eqnarray}
&&(\Phi_{\alpha-2}^{*\gamma+2},V_{\text{I}}
\Phi_{\alpha}^{\gamma}) = \nonumber \\
&&\frac{1}{2}\sqrt{n_{\alpha}(n_{\alpha}-1)(n_{\gamma}+1)(n_{\gamma}+2)}
\langle \gamma \gamma|v_{12}|\alpha \alpha \rangle \nonumber \\
\label{case4}
\end{eqnarray} 

\subsection{Conditional probabilty distributions}
If the many-body wave function is denoted as 
$\Psi({\bf r}_{1},{\bf r}_{2},\ldots,{\bf r}_{N})$, 
the conditional probability of finding a particle at position ${\bf r}$ 
given that there is another one at position ${\bf r}_0$ is given by
\begin{equation}
P({\bf r},{\bf r}_0)=\frac{N(N-1)}{2} \int \prod_{i=3,N}d{\bf r}_{i}
|\Psi({\bf r},{\bf r }_{0};{\bf r }_{3}, \ldots,{\bf r}_{N})|^2.
\end{equation} 
We reduce the calculation of the conditional probability (within a 
proportionality constant) to evaluating the 
expectation value of the following two-body operator
\begin{equation}
T=\sum_{i \neq j}\delta({\bf r}-{\bf r}_{i})\delta({\bf r}_0-{\bf r}_{j}).
\end{equation}

If ${\bf r} \neq {\bf r}_0$, the expression under the sum is not symmetric with 
respect to interchanging ${\bf r}_{i}$ and ${\bf r}_{j}$ [unlike the case
of the two-body potential $v({\bf r}_i -{\bf r}_j)$]. We rewrite $T$ as:
\begin{eqnarray}
T&=&\sum_{i < j}\delta({\bf r}-{\bf r}_{i})\delta({\bf r}_0-{\bf r}_{j})+
\sum_{j < i} \delta({\bf r}-{\bf r}_{i})\delta({\bf r}_0-{\bf r}_{j}) 
\nonumber \\
&=&\sum_{i < j} \biglb( \delta({\bf r}-{\bf r}_{i})\delta({\bf r}_0-{\bf r}_{j})+
\delta({\bf r}-{\bf r}_{j})\delta({\bf r}_0-{\bf r}_{i}) \bigrb). \nonumber \\
\label{t_rewrite}
\end{eqnarray}
which has a {\em symmetric} expression under the summation. 
Then the calculation of the CPD proceeds exactly as the calculation for the
interaction part of the Hamiltonian.


\begin{thebibliography}{99}
\bibitem{stri}
F. Dalfovo, S. Giorgini, L.P. Pitaevskii, and S. Stringari,
Rev. Mod. Phys. {\bf 71}, 463 (1999).
\bibitem{legg}
A. J. Leggett,
Rev. Mod. Phys. {\bf 73}, 307 (2001).
\bibitem{ho}
T.L. Ho, 
Phys. Rev. Lett. {\bf 87}, 060403 (2001).
\bibitem{fett}
A.L. Fetter,
Phys. Rev. A {\bf 64}, 063608 (2001).
\bibitem{peth}
C.J. Pethick and H. Smith,
{\it Bose-Einstein Condensation in Dilute Gases\/}
(Cambridge University Press, Cambridge, 2002).
\bibitem{pita}
L.P. Pitaevskii and S. Stringari, 
{\it Bose-Einstein Condensation\/}
(Oxford University Press, Oxford, 2003).
\bibitem{baks1}
L.O. Baksmaty, S.J. Woo, S. Choi, and N. P. Bigelow, 
Phys. Rev. Lett. {\bf 92}, 160405 (2004).
\bibitem{baks2}
S.J. Woo, L.O. Baksmaty, S. Choi, and N. P. Bigelow, 
Phys. Rev. Lett. {\bf 92}, 170402 (2004). 
\bibitem{peth2}
G. Watanabe, G. Baym, and C.J. Pethick,
Phys. Rev. Lett. {\bf 93}, 190401 (2004).
\bibitem{baym1}
G. Baym,
J. Low Temp. Phys. {\bf 138}, 601 (2005).
\bibitem{dali}
K.W. Madison, F. Chevy, W. Wohlleben, and J. Dalibard,
Phys. Rev. Lett. {\bf 84}, 806 (2000).
\bibitem{kett}
C. Raman, J.R. Abo-Shaeer, J.M. Vogels, K. Xu, and W. Ketterle,
Phys. Rev. Lett. {\bf 87}, 210402 (2001).
\bibitem{corn}
P. Engels, I. Coddington, P.C. Haljan, and E.A. Cornell,
Phys. Rev. Lett. {\bf 89}, 100403 (2002).
\bibitem{wilk1} 
N.K. Wilkin and J.M.F. Gunn, 
Phys. Rev. Lett. {\bf 84}, 6 (2000).
\bibitem{vief}
S. Viefers, T.H. Hansson, and S.M. Reimann,
Phys. Rev. A {\bf 62}, 053604 (2000).
\bibitem{wilk2} 
N.R. Cooper, N.K. Wilkin, J.M.F. Gunn, 
Phys. Rev. Lett. {\bf 87}, 120405 (2001).
\bibitem{regn1}
N. Regnault and Th. Jolicoeur, 
Phys. Rev. Lett. {\bf 91}, 030402 (2003).
\bibitem{regn2} 
C.C. Chang, N. Regnault, Th. Jolicoeur, and J.K. Jain, 
Phys. Rev. A {\bf 72}, 013611 (2005).
\bibitem{yl1}
C. Yannouleas and U. Landman,
Phys. Rev. B {\bf 66}, 115315 (2002). 
\bibitem{yl2}
C. Yannouleas and U. Landman,
Phys. Rev. B {\bf 68}, 035326 (2003). 
\bibitem{yl3}
C. Yannouleas and U. Landman,
Phys. Rev. B {\bf 69}, 113306 (2004).
\bibitem{yl4}
C. Yannouleas and U. Landman,
Phys. Rev. B {\bf 70}, 235319 (2004).
\bibitem{yl5}
Y.S. Li, C. Yannouleas, and U. Landman,
Phys. Rev. B {\bf 73}, 075301 (2006).
\bibitem{yl6}
C. Yannouleas and U. Landman,
phys. stat. sol. (a) {\bf 203}, 1160 (2006).
\bibitem{bao}
W.Y. Ruan, Y.Y. Liu, C.G. Bao, and Z.Q. Zhang,
Phys. Rev. B {\bf 51}, 7942 (1995). 
\bibitem{seki}
T. Seki, Y. Kuramoto, and T. Nishino,
J. Phys. Soc. Jpn. {\bf 65}, 3945 (1996).
\bibitem{maks}
P.A. Maksym,
Phys. Rev. B {\bf 53}, 10 871 (1996).
\bibitem{yl7}
C. Yannouleas and U. Landman,
Phys. Rev. Lett. {\bf 82}, 5325 (1999).
\bibitem{egge}
R. Egger, W. H\"{a}usler, C.H. Mak, and H. Grabert,
Phys. Rev. Lett. {\bf 82}, 3320 (1999).
\bibitem{yl8}
C. Yannouleas and U. Landman,
Phys. Rev. Lett. {\bf  85}, 1726 (2000).
\bibitem{mikh}
S.A. Mikhailov,
Phys. Rev. B {\bf 65}, 115312 (2002).
\bibitem{yl9}
C. Yannouleas and U. Landman,
Phys. Rev. B {\bf 68}, 035325 (2003). 
\bibitem{yl10}
C. Ellenberger, T. Ihn, C. Yannouleas, U. Landman, K. Ensslin, D. Driscoll, 
and A.C. Gossard,
Phys. Rev. Lett. {\bf 96}, 126806 (2006).
\bibitem{yl11}
I. Romanovsky, C. Yannouleas and U. Landman,
Phys. Rev. Lett. {\bf 93}, 230405 (2004).
\bibitem{yl12}
C. Yannouleas and U. Landman,
Proc. Natl. Acad. Sci. (USA) {\bf 103}, 10600 (2006).
\bibitem{yl13}
I. Romanovsky, C. Yannouleas, L.O. Baksmaty, and U. Landman,
Phys. Rev. Lett. {\bf 97}, 090401 (2006).
\bibitem{mann}
S.M. Reimann, M. Koskinen, Y. Yu, and M. Manninen, 
New J. Phys. {\bf 8}, 59 (2006).
\bibitem{barb}
N. Barberan, M. Lewenstein, K. Osterloh, and D. Dagnino, 
Phys. Rev. A {\bf 73}, 063623 (2006). 
\bibitem{note1}
The use of analytic variational (referred to also as trial) many-body wave 
functions restricted to the LLL has a long history starting with the Laughlin
wave functions [see R.B. Laughlin, Phys. Rev. Lett. {\bf 50}, 1395 (1983)]. 
For the derivation of analytic rotating-electron-molecule 
functions restricted to the LLL, see Ref.\ \cite{yl1}. For the 
generalization of the REM variational functions beyond the LLL, see Refs.\ 
\cite{yl3,yl5}.
\bibitem{note56}
For the definition of the fractional filling $\nu$, we follow 
the fractional quantum Hall effect literature, see, e.g., 
S.M. Girvin and T. Sach, Phys. Rev. B {\bf 28}, 4506 (1983).
\bibitem{orsay}
M. Shellekens, R. Hoppeler, A. Perrin, J.V. Gomes, D. Boiron, 
A. Aspect, and C.I. Westbrook, 
Science {\bf 310}, 648 (2005).
\bibitem{shot_noise_1}
E. Altman, E. Demler, and M. D. Lukin, 
Phys. Rev. A {\bf 70}, 013603 (2004).
\bibitem{shot_noise_2}
S. F\"{o}lling, F. Gerber, A. Widera, O. Mandel, T. Gericke, and I. Bloch, 
Nature (London) {\bf 434}, 481 (2005).
\bibitem{note2}
Ref.\ \cite{mann} used a Gaussian-type two-body interaction in place of a
contact potential.
\bibitem{note67}
EXD calculations for $N=6$ {\it electrons\/} in the LLL were also performed  
in M. Manninen, S. Viefers, M. Koskinen, and S.M. Reimann, Phys. Rev. B 
{\bf 64}, 245322 (2001).
These calculations were made in the same angular-momentum range ($L \leq 60$) 
as the earlier Refs.\ \cite{seki,maks}. For extensive EXD calculations for $N=6$
electrons [covering the full range $15 \leq L \leq 135$ 
$(1 \geq \nu \geq 1/9)$], see Ref.\ 
\cite{yl2}; such an extensive range is necessary for demonstrating the 
superiority of the REM interpretation for QDs under high $B$ compared to the 
composite-fermion and Jastrow-Laughlin ``special-quantum-liquid'' one (for the 
latter interpretation in the limited range $1 \geq \nu \geq 1/3$, see J.K. 
Jain and T. Kawamura, Europhys. Lett. {\bf 29}, 321 (1995), and also Refs.\ 
\cite{seki,maks}).
\bibitem{feder} 
Alex G. Morris and David L. Feder, e-print cond-mat/0602037. 
\bibitem{fock}
V. Fock,
Z. Phys. {\bf 47}, 446 (1928).
\bibitem{darw}
C.G. Darwin,
Proc. Cambridge Philos. Soc. {\bf 27}, 86 (1930).
\bibitem{arp}
R. B. Lehoucq, D. C. Sorensen, and C. Yang, 
{\it ARPACK Users' Guide: Solution of Large-Scale Eigenvalue Problems with 
Implicitly Restarted Arnoldi Methods\/} (SIAM, Philadelphia, 1998).
\bibitem{petsc}
ARPACK was used in conjunction with PETSc (http://www.mcs.anl.gov/petsc)
and SLEPc (http://www.grycap.upv.es/slepc).
\bibitem{note3}
For spin-polarized electrons in the LLL, the corresponding value is 
$L_0=N(N-1)/2$ [see Refs.\ \cite{yl2,yl5}].
\bibitem{peeters_classical} 
M. Kong, B. Partoens, and F.M. Peeters, 
Phys. Rev. E {\bf 65}, 046602 (2002).
\bibitem{macd}
A.H. MacDonald, S.R.E. Yang, and M.D. Johnson,
Aust. J. Phys. {\bf 46}, 345 (1993).
\bibitem{youl}
S. Yi and L. You,
Phys. Rev. Lett. {\bf 92}, 193201 (2004). 
\bibitem{note57}
A single particle at the center does not count as an independent ring
towards this rule.
\end{thebibliography}
\end{document}